\documentclass[8.5pt,twoside,twocolumn]{article}
\pdfoutput=1
\usepackage[super,sort&compress,comma]{natbib} 
\usepackage[version=3]{mhchem}
\usepackage{balance}
\usepackage{times}
\usepackage{sectsty}
\usepackage[pdftex]{graphicx}% Include figure files
\usepackage{amsmath}% Include figure files
\usepackage{amssymb}
\usepackage{lastpage}
\usepackage[format=plain,justification=raggedright,singlelinecheck=false,font=small,labelfont=bf,labelsep=space]{caption} 
\usepackage{fancyhdr}
\usepackage[dvips,usenames]{color}
\usepackage{placeins}
\usepackage{bm}
\usepackage{pdfpages}

\begin{document}

\thispagestyle{plain}
\fancypagestyle{plain}{
\fancyhead[L]{\includegraphics[height=8pt]{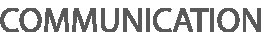}}
\fancyhead[C]{\hspace{-1cm}\includegraphics[height=20pt]{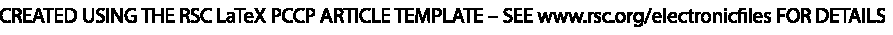}}
\fancyhead[R]{\hspace{10cm}\vspace{-0.25cm}\includegraphics[height=10pt]{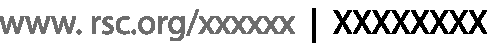}}
\renewcommand{\headrulewidth}{1pt}}
\renewcommand{\thefootnote}{\fnsymbol{footnote}}
\renewcommand\footnoterule{\vspace*{1pt}% 
\hrule width 3.4in height 0.4pt \vspace*{5pt}} 
\setcounter{secnumdepth}{5}

\makeatletter 
\renewcommand\@biblabel[1]{#1}            
\renewcommand\@makefntext[1]% 
{\noindent\makebox[0pt][r]{\@thefnmark\,}#1}
\makeatother 
\renewcommand{\figurename}{\small{Fig.}~}
\sectionfont{\large}
\subsectionfont{\normalsize} 

\fancyfoot{}
\fancyfoot[LO,RE]{\vspace{-7pt}\includegraphics[height=9pt]{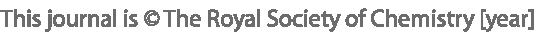}}
\fancyfoot[CO]{\vspace{-7.2pt}\hspace{12.2cm}\includegraphics{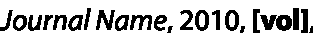}}
\fancyfoot[CE]{\vspace{-7.5pt}\hspace{-13.5cm}\includegraphics{RF}}
\fancyfoot[RO]{\footnotesize{\sffamily{1--\pageref{LastPage} ~\textbar  \hspace{2pt}\thepage}}}
\fancyfoot[LE]{\footnotesize{\sffamily{\thepage~\textbar\hspace{3.45cm} 1--\pageref{LastPage}}}}
\fancyhead{}
\renewcommand{\headrulewidth}{1pt} 
\renewcommand{\footrulewidth}{1pt}
\setlength{\arrayrulewidth}{1pt}
\setlength{\columnsep}{6.5mm}
\setlength\bibsep{1pt}

\twocolumn[
  \begin{@twocolumnfalse}
\noindent\LARGE{\textbf{Self-propulsion of a catalytically active particle near a planar wall: from
reflection to sliding and hovering$^\dag$}}
\vspace{0.6cm}

% \textit{$^{a\ddag}$}
\noindent\large{\textbf{W.~E. Uspal,\textit{$^{ab}$} M.~N. Popescu,\textit{$^{abc}$} S. Dietrich,\textit{$^{ab}$} and
M. Tasinkevych$^{\ast}$\textit{$^{ab}$}}}\vspace{0.5cm}
%Please note that \ast indicates the corresponding author(s) but no footnote text is required. 

\noindent\textit{\small{\textbf{Received Xth XXXXXXXXXX 20XX, Accepted Xth XXXXXXXXX 20XX\newline
First published on the web Xth XXXXXXXXXX 200X}}}

\noindent \textbf{\small{DOI: 10.1039/b000000x}}
 \end{@twocolumnfalse} \vspace{0.6cm}

  ]

\noindent\textbf{Micron-sized particles moving through solution in response to self-generated chemical
gradients serve as model systems for studying active matter. Their far-reaching potential
applications will require the particles to sense and respond to their local environment in
a robust manner. The self-generated hydrodynamic and chemical fields, which induce
particle motion, probe and are modified by that very environment, including confining
boundaries. Focusing on a catalytically active Janus particle as a paradigmatic example,
we predict that near a hard planar wall such a particle exhibits several scenarios of
motion: reflection from the wall, motion at a steady-state orientation and
height above the wall, or motionless, steady ``hovering.'' Concerning the steady states,
the  height and the orientation are determined both by the proportion of
catalyst coverage and the interactions of the solutes with the different ``faces'' of
the particle. Accordingly, we propose that a desired behavior can be selected by tuning
these parameters via a judicious design of the particle surface chemistry.}
\section*{}
\vspace{-1cm}
%Footnotes
%\footnotetext{\dag~Electronic Supplementary Information (ESI) available: Further details concerning the %numerical method and analytical and numerical results. See DOI: 10.1039/b000000x/}

%Please use \dag to cite the ESI in the main text of the article.
%If you article does not have ESI please remove the the \dag symbol from the title and the above footnotetext.

\footnotetext{\textit{$^{a}$~Max-Planck-Institut f\"{u}r Intelligente Systeme, Heisenbergstr. 3, D-70569
Stuttgart, Germany.}}
\footnotetext{\textit{$^{b}$~IV. Institut f\"ur Theoretische Physik, Universit\"{a}t Stuttgart,
Pfaffenwaldring 57, D-70569 Stuttgart, Germany.}}
\footnotetext{\textit{$^{c}$~Ian Wark Research Institute, University of South Australia, Adelaide, SA
5095, Australia.}}

%additional addresses can be cited as above using the lower-case letters, c, d, e... If all authors are from the same address, no letter is required

%\footnotetext{\ddag~Additional footnotes to the title and authors can be included \emph{e.g.}\ `Present address:' or `These authors contributed equally to this work' %as above using the symbols: \ddag, \textsection, and \P. Please place the appropriate symbol next to the author's name and include a \texttt{\textbackslash %footnotetext} entry in the the correct place in the list.}

%\footnotetext{*To whom correspondence should be addressed: }

Autonomous microscopic agents moving through confined, liquid-filled spaces are envisioned as a key component of future lab-on-a-chip and drug delivery systems.\cite{patra12} Chemically active Janus particles offer a realization of such agents. A Janus ``micromotor'' works by catalytically activating, over a fraction of its surface, chemical reactions in the surrounding solution. \textcolor{black}{The resulting chemical gradients can drive directed motion through a variety of mechanisms: self-electrophoresis, in which ionic currents drive the motion, of bi-metallic particles;$^{\ddag}$\cite{paxton04,paxton06b} \footnotetext{\ddag~\textcolor{black}{Recently it was argued that self-electrophoresis may also be a possible mechanism for silica particles covered with platinum.\cite{brown14,ebbens14}}}bubble propulsion,
in particular for active tubes covered on the inside by a catalyst;\cite{gao12} and self-diffusiophoresis, in which the reaction product is electrically neutral, such as for silica or polystyrene spheres covered by platinum.\cite{golestanian05,golestanian07,howse07,baraban12} Recent reviews catalog and detail these and other propulsion mechanisms.\cite{ebbens10, poon13, wang13}} Janus micromotors have been \textcolor{black}{harnessed} for
applications such as transportation of inert cargo \cite{baraban12} and environmental remediation.\cite{gao13}
% *** change ***
% removal of environmental contaminants

%%%%%%%%%%%%%%%%%%%%%%%%%%%%%%%%%%
\begin{figure}[th]
\centering
\includegraphics[width=0.95 \columnwidth]{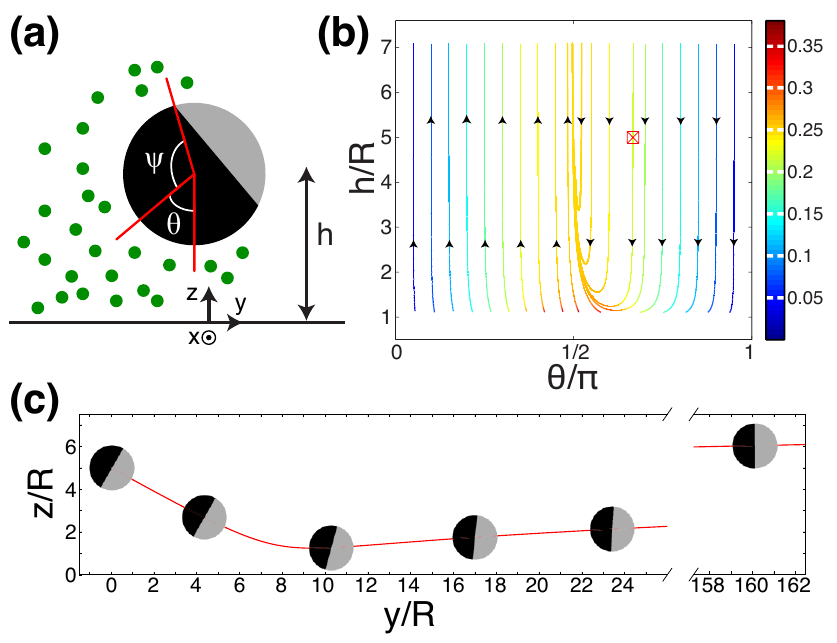}
\caption{
\label{fig:reflecting_chi0_0.0} (a) Schematic diagram of the model system.  A sphere of
radius $R$ has a surface which is partially catalytic (black) and partially inert (grey). 
The extent of the catalytic cap is parametrized by $\chi_{0} \equiv -\cos(\psi)$. 
%Over the catalytic part, a chemical product (green discs) is produced  and diffuses in the
%solution.
Green discs \textcolor{black}{indicate} a diffusing chemical product. The system is bounded by an inert wall.
The height $h$ and the orientation angle $\theta$ specify the 
configuration of the sphere.
%The configuration of the Janus sphere is specified by the height $h$ of its center
%above the wall and by the orientation angle $\theta$ of the symmetry axis
%f the Janus sphere relative to the wall normal.  
(b) Phase plane for half coverage by catalyst ($\chi_{0} = 0$) of the Janus particle, where
the color encodes $U_{y}/U_{0}$.
%, i.e., the component of the normalized velocity of the particle parallel to the wall.
(c) A typical trajectory of such a Janus particle.
% Upon moving away from the wall,
% *** change ***
%(right side of the panel)
%the orientation of the particle attains a value $\theta
%\lesssim \pi/2$.
 The initial configuration  % of the particle
is $h_{0}/R = 5$ and $\theta_{0} = 120^{\circ}$, which is indicated in (b) 
with the symbol \textcolor{red}{$\boxtimes$}.
}
\end{figure}
%%%%%%%%%%%%%%%%%%%%%%%%%%%%%%%55

Recently, several studies have sought to isolate and understand the role of confinement
in determining the particle motion.  The behavior upon collisions with the confining
boundaries was explored in experiments using particles moving in microchannels. 
Significant motion of Janus particles along the microchannel walls was observed, with
subsequent detachment \textcolor{black}{attributed to} reorientation of the particle due to thermal noise.\cite{volpe11, kreuter13} \textcolor{black}{Bi-metallic swimming rods have been observed to orbit around stationary spherical colloids. This behavior has been semi-quantitatively captured via lubrication analysis.\cite{takagi14}} In two dimensions, the scenarios of a particle hitting or escaping from a wall are captured by the ``Janus active disc'' model analyzed in Ref. \citenum{crowdy13}. For certain geometrical configurations and model Janus particles, such as
a spherical particle in the center of a spherical cavity \cite{popescu09} or a dimer
translating along the axis of a square tube in a Poiseuille flow,\cite{tao10} the
dependence of the particle velocity on the characteristic size of the confinement was
obtained via analytical or numerical calculations.  For the case of ``mechanical
swimmers,''  modeling micro-organisms which move via shape changes, it was shown
theoretically that when motion occurs near a boundary, hydrodynamic interactions can
induce a rich dynamical behavior similar to that observed for bacteria\cite{lauga06,lauga08,spagnolie12,ishimoto13} \textcolor{black}{and robotic swimmers.\cite{zhang10}}

Here, we investigate  self-diffusiophoresis of a catalytically active spherical Janus
particle near a planar boundary.  The particle ``senses'' and responds to the presence of
the boundary via the chemical and hydrodynamic fields it creates.  Chemically, the
particle effectively releases a solute from a catalytic region of its surface. \textcolor{black}{The resulting anisotropic distribution of
solute drives a surface flow in a thin layer surrounding the particle,
leading to directed motion.\cite{anderson89, golestanian05,
golestanian07, popescu10, poon13}} The wall is impenetrable to the solute, modifying the solute number density at the particle
surface. Hydrodynamically, the particle creates disturbance flows in the fluid, and these
flows are reflected by the no-slip boundary, coupling back to the particle.  The issue is
to understand how this relation between sensing and response depends on the surface
chemistry of the particle.

%%%%%%%%%%%%%%%%%%%%%%%%%%%%%%%%%%%%%%%%%%%%%
\begin{figure*}[h!tb]
\centering
\includegraphics[width=0.9\linewidth]{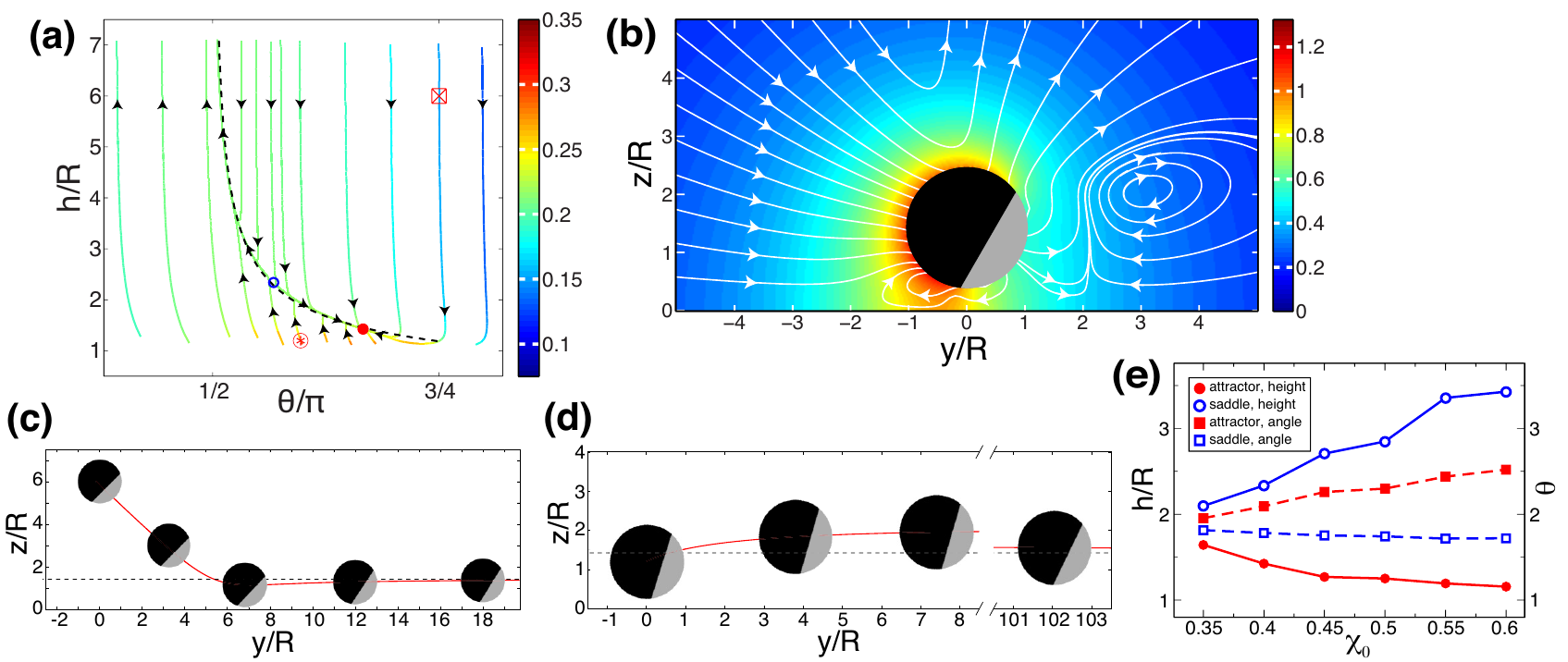}
\caption{\label{fig:reflecting_chi0_0.4} (a) Phase plane for $\chi_0 = 0.4$. Colors
indicate $U_{y}/U_{0}$.  There is a dynamical attractor at $h_{eq}/R = 1.42$ and
$\theta_{eq} = 119.9^{\circ}$ (solid red circle).
 % the particle moves along the wall with constant $h$ and $\theta$. 
The dashed curve shows the analytically estimated ``slow manifold'' $U_{z} = 0$ (see main text) with a
numerically fitted prefactor. 
%In the regions of the phase space $\theta < \pi/4$ and
%$\theta > 4 \pi/5$ not shown here, the particle either escapes from or ``crashes'' into
%the wall. 
(b) Flow field in the laboratory frame (white streamlines) and the solute number density $c/c_{0}$ (colors) associated with the sliding
state. (c) A typical trajectory with the initial configuration $h_{0}/R = 6$ and $\theta_{0} =  3 \pi/4$ in the basin of attraction
for sliding (\textcolor{red}{$\boxtimes$} in (a)). (d) A trajectory with $h_{0}/R = 1.2$ and
$\theta_{0} = 107.5^{\circ}$ (\textcolor{red}{$\circledast$} in (a)). (e) Variation of the
locations of the attractor (\textcolor{red}{$\bullet$}) and of the saddle point
(\textcolor{blue}{$\circ$}) with $\chi_{0}$.}
\end{figure*}
%%%%%%%%%%%%%%%%%%%%%%%%%%%%%%%%%%%%%%%%%%%%%%%%%%%%

Here, we demonstrate that qualitatively distinct dynamics can
be evoked by varying certain particle design parameters:  (i) the proportion of catalyst
coverage, (ii) the repulsive or attractive character of the solute-particle interactions,
and (iii) the relative strength of the interactions of the solute with the catalytic and
inert particle faces. In particular, for high catalyst coverage and identical repulsive
interactions a particle attains a stable state in which it slides along the wall at a fixed height and
orientation. Similar dynamics may be obtained for moderate catalyst coverage, but
stronger repulsion of the solute from the catalytic than from the inert face.
For very high catalyst coverage and repulsive interactions, a particle
attains a stable hovering state in which it acts as a stationary micropump.  We develop
simple quantitative models which shed light on the physical mechanisms sustaining these
steady states. We anticipate that these findings can be used in microfluidic devices to
create robust and predictable motion of active particles either away from or near walls.
% *** change ***
%as desired in specific applications. 

%%%%%%%%%%%%%%%%%%%%%%%%%%%%%%%%%%%%%%%%%%
%\begin{figure*}[tbh!]
\begin{figure*}[h!tb]
\centering
\includegraphics[width=0.9\linewidth]{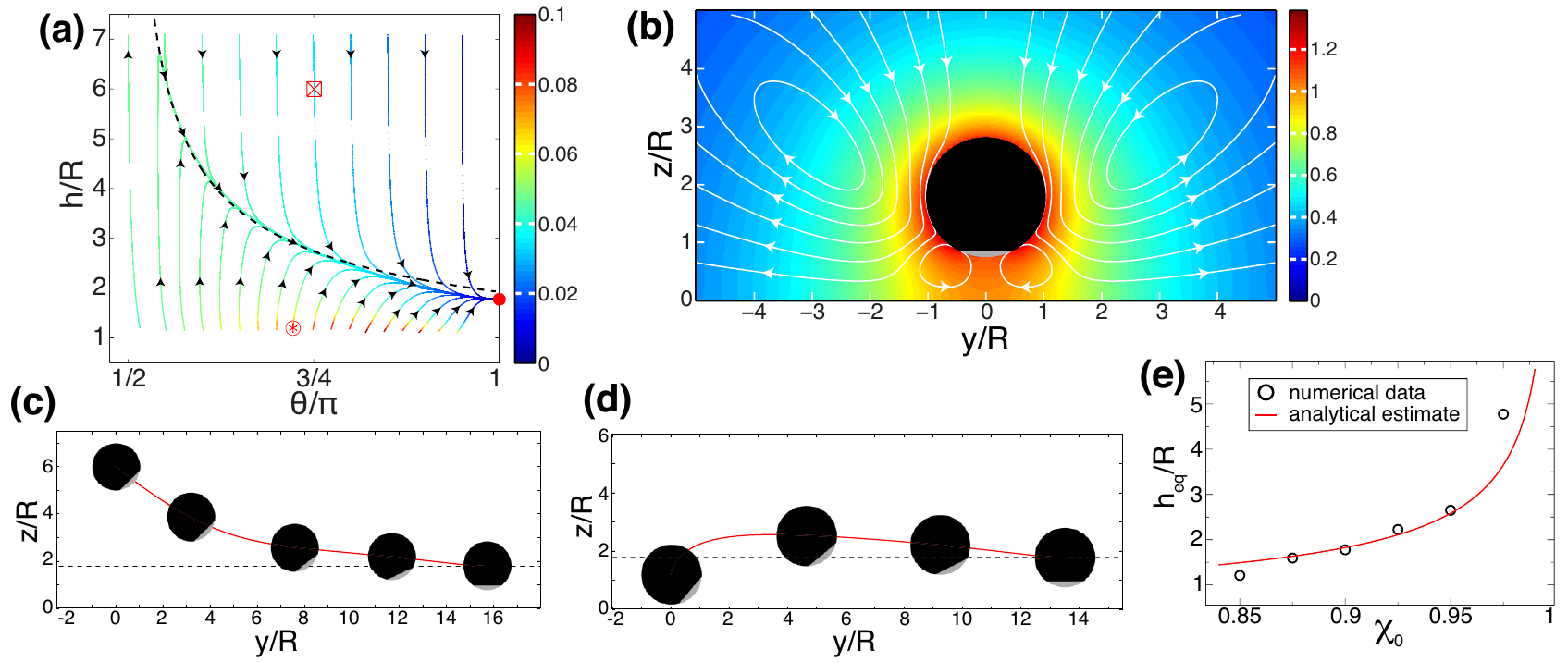}
\caption{\label{fig:reflecting_chi0_0.9} (a) Phase plane for $\chi_0 = 0.9$. Colors
indicate $U_{y}/U_{0}$.  There is an attractor at $h_{eq}/R = 1.78$ and $\theta_{eq} =
180^{\circ}$ (solid red circle).  This is a ``hovering'' state in which the particle
remains fixed in space. The dashed curve shows the fitted slow manifold.
%In the section $\theta < \pi/2$ of the phase space
%(not shown) the particle escapes. 
(b) Flow field (white streamlines) and solute
number density $c/c_{0}$ (colors) for the hovering state. These fields
are rotationally symmetric about the axis $x = 0$, $y = 0$.  (c) An exemplary trajectory
with initial configuration $h_{0}/R = 6$ and $\theta_{0} = 3\pi/4$
\textcolor{red}{$\boxtimes$} in (a)).
% *** change ***
% which belongs to the basin of attraction for hovering.
(d) The trajectory for $h_{0}/R = 1.2$ and $\theta_{0} = 130^{\circ}$ (\textcolor{red}{$\circledast$} in
(a)). (e) Variation of the height $h_{eq}/R$ of the hovering state with catalyst coverage
$\chi_{0}$.  
%The numerical data broadly agree with an analytical estimate obtained by
%modeling the effect of the wall in terms of an image source monopole.
}
\end{figure*}
%%%%%%%%%%%%%%%%%%%%%%%%%%%%%%%%%%

As shown in Fig.~\ref{fig:reflecting_chi0_0.0}(a), we consider a spherical particle of
radius $R$. The particle is covered by a catalyst over a spherical cap region 
parametrized by $\chi_{0} = -\cos(\psi)$ (black segment in Fig.~\ref{fig:reflecting_chi0_0.0}). 
%The ``polarity'' of the particle is defined by the diameter passing through the center of the catalytic cap and oriented
%towards the catalyst. %The extent of the catalytic cap is parametrized by $\chi_{0} = \cos(\psi)$. 
The sphere is suspended in a Newtonian liquid solution bounded by a 
chemically inert  planar wall located at $z = 0$. The catalytic cap releases a solute
which diffuses in the solution. The wall is impenetrable to the solute and
there are no other solute-wall interactions. There are effective interactions between the
solute and the particle surface. Due to the symmetry of the
system and in the absence of thermal fluctuations, the particle moves only in the plane
containing the wall normal and the particle's symmetry axis. Therefore, the cap orientation $\theta$ 
and the height $h$ of the particle's center above the wall completely specify the particle configuration.
 The translational and angular velocities are denoted by $\mathbf{U}$ and $\mathbf{\Omega}$, respectively. We
consider the motion of the sphere to be sufficiently slow and the diffusion of the solute
to be sufficiently fast such that at each instantaneous $(h,\theta)$ a quasi-steady state of the solute number density
$c(\mathbf{r})$ and of the hydrodynamic flow $\mathbf{u}(\mathbf{r})$ is
established.
%i.e., $\mathbf{u}(\mathbf{r})$ and $c(\mathbf{r})$ depend on time only through the time
%dependence of $h$ and $\theta$.

We calculate the self-propulsion velocities $\mathbf{U}$ and $\mathbf{\Omega}$ by
employing the classical theory of diffusiophoresis,\cite{derjaguin:47,anderson89} which
is briefly discussed in the {\it Supplemental Information} ({\it SI}) along with
the associated numerical approach.
%
%As we discuss in {\it SM}, the reciprocal theorem
%implies that the problem of finding $\mathbf{V} \equiv (\mathbf{U},\mathbf{\Omega})$
%reduces to the solution of a system of six linear equations.  
%
$\mathbf{U}$ and $\mathbf{\Omega}$ are calculated over a grid of $\theta$ and $h$, which  
is limited to $h/R \geq 1.1$ for the sake of numerical accuracy and for ensuring the validity of the quasi-steady
state approximations discussed above. In order to obtain a full particle trajectory for a
certain initial condition $(h_{0}, \theta_{0})$, we perform numerical integration by
interpolating $\dot{h} = U_{z}$, $\dot{y} = U_{y}$, and $\dot{\theta} = -\Omega_{x}$ from
the grid. 
%Because of the symmetry of the problem and the absence of rotational
%diffusion, there is no rotation around the $y$ or $z$ axes. 
% *** change ***
% Note that because of the symmetry of the problem and the absence of rotational
%diffusion, there is no rotation around the $y$ or $z$ axes, i.e., $\Omega_{y} = \Omega_{z}
%= 0$.
In the following, the solute number density will be expressed in units of $c_{0}
\equiv | \kappa R/D |$ and the velocity in units of $U_{0} \equiv |b \kappa/D|$. $D$ is
the diffusion coefficient of the solute and $\kappa$ is the rate of solute production
per area at the cap; $b$ is a ``surface mobility''; its magnitude and sign
reflect the strength and the attractive or repulsive character of the
interaction between the solute and the particle surface.\cite{anderson89}

We first consider a half-covered sphere ($\chi_{0} = 0$),
%,which is the most common chemical configuration used in experiments.  
and assume uniform repulsion $(b<0)$ of the solute from the particle surface.
% (i.e., $b$ is uniform over the particle surface.) 
Our results are summarized in the phase plane of Fig.~\ref{fig:reflecting_chi0_0.0}(b).
%, which shows the evolution of a configuration $(h_0, \theta_0)$. 
Trajectories starting from initial orientations $\theta_0 \leq \pi/2$ move away from the wall without a
significant change of $\theta$. For $\theta_0 > \pi/2$ the particle exhibits a richer
dynamics.  \textcolor{black}{A representative trajectory, with the initial condition given by the symbol \textcolor{red}{$\boxtimes$} in the phase plane [Fig.~\ref{fig:reflecting_chi0_0.0}(b)], is shown in Fig.~\ref{fig:reflecting_chi0_0.0}(c). The particle moves towards the wall, approximately maintaining its initial orientation of $\theta_{0} = 120^{\circ}$ until its scaled height is less than $h/R \simeq 1.5$. In close vicinity of the wall, the particle rotates its catalytic cap towards the wall, e.g. to approximately 100$^\circ$ at $y/R = 10$, and further towards $\gtrsim 90^\circ$ at $y/R =
17$.  It escapes from the wall with an asymptotic orientation $\theta \lesssim 90^{\circ}$, which is independent of the initial condition.}
%Initially, the particle moves towards the wall with
%little rotation. In close vicinity of the wall, the particle reorients its
%catalytic cap towards the wall and escape from the wall, with the asymptotic orientation
%$\theta \lesssim \pi/2$ independent of the initial condition. 
This behavior -- reflection from the wall -- is similar to the results derived in Ref. \citenum{crowdy13} for
the motion of a half emitting, half absorbing disc, as well as to the results in
Ref. \citenum{spagnolie12} for a spherical ``squirmer'' near a wall.  
Note that for larger  $\theta_0$ the turning point is located closer to the wall;
thus, many of the trajectories in this region of the phase
space appear to ``crash'' into the wall (Fig. \ref{fig:reflecting_chi0_0.0}(b)). 
This is, however, just a numerical artifact meaning that the turning point of the trajectory 
is below the minimum allowed height. 
%It seems
%to be a reasonable expectation that this is just a numerical artifact, i.e., that an
%apparent ``crash'' simply means that the turning point of the trajectory is below the
%minimum height which we have allowed in the numerical solution. 

% location 2

The mechanism behind the turning point must be driven by the wall through two possible
effects: either via the wall induced changes in the solute gradients,
affecting the phoretic slip, or via the confinement of the hydrodynamic flow.
Smoluchowksi found that solute gradients cannot drive rotation of a particle with uniform $b$,\cite{anderson89}
hence we anticipate that rotation is dominated by hydrodynamic interaction. 
The hydrodynamic and chemical effects can be identified and isolated by performing the
numerical calculations with properly chosen boundary  conditions (see {\it SI}). 
%By using this approach, we have been able to confirm  that the rotation is solely due 
%to the disturbance flows reflected  by the wall (see {\it SM}).
\textcolor{black}{By using this approach, we have been able to confirm that chemical contributions to the rotation of the particle are negligible, except potentially very close to the wall (see {\it SI}).}

%The mechanism behind the turning point can be understood as follows. Since the sphere
%does not rotate if unconfined, rotation must be driven by the wall through two possible
%effects: either via the wall induced changes in the solute gradients along the surface of
%the particle, affecting the phoretic slip, or via the confinement of the hydrodynamic flow
%between the particle and the wall.  Smoluchowksi found that solute gradients cannot drive
%rotation of a particle which has uniform surface mobility \cite{anderson89}. As we
%assumed uniform surface mobility, we anticipate that any rotation is strictly
%hydrodynamic in origin. The hydrodynamic and chemical effects can be identified and
%isolated by performing the numerical calculations with different boundary conditions.
%Calculating the particle velocity under the assumption of an unconfined solute
%number density $c(\mathbf{r})$ (but a confined flow field $\mathbf{u} (\mathbf{r})$)
%isolates the hydrodynamic effect of the wall.  The chemical effect is isolated by
%calculating the particle velocity on the basis of $\mathbf{u}(\mathbf{r})$ being
%unconfined (but $c(\mathbf{r})$ being confined). By using this approach, we have been
%able to confirm that the rotation is solely due to the disturbance flows reflected by the wall (see {\it SM}).

% *** change ***
%i.e., the turning point, 

In all cases the rotation of the particle is such as to favor subsequent motion away
from the wall, i.e., $\dot{\theta} < 0$ for all angles $0 < \theta < \pi$. Thus
the question arises if states with $\dot{\theta} > 0$, i.e., turning towards the wall, or
states with $\dot{\theta} = 0$ and $U_{z} = 0$, which would represent a dynamical fixed point 
%corresponding to motion at a fixed height and orientation,
do exist. To address this question, we note that the disturbance flows created by the particle can be described via
a superposition of ``hydrodynamic singularities,'' which are terms in a multipole expansion, centered on the particle, for the flow field surrounding it.
The strengths of these singularities determine if and where a curve in the phase
plane with $\dot{\theta} = 0$ exists.\cite{spagnolie12} The singularity strengths
can be tuned by varying the coverage of the particle by catalyst. For coverages in the
range $-1 < \chi_{0} < 0.35$, the phase planes and trajectories resemble those obtained for half coverage (see 
Fig. \ref{fig:reflecting_chi0_0.0}(b)). However, at $\chi_{0} = 0.35$ a bifurcation
occurs: a saddle point and a dynamical attractor emerge. These are illustrated in Fig.
\ref{fig:reflecting_chi0_0.4}(a), which shows a phase plane for $\chi_{0} = 0.4$. Notably,
many trajectories converge, without overlap, to a single curve indicated by the
dashed line, which includes both the attractor ($(h_{eq}, \theta_{eq})$, solid
red circle) and the saddle point (open blue circle). The attractor represents a
``sliding'' state: the particle maintains a fixed height and orientation as it moves
along the wall. The structure of $\mathbf{u}(\mathbf{r})$ and $c(\mathbf{r})$
corresponding to the sliding state is shown in Fig.~\ref{fig:reflecting_chi0_0.4}(b).
%For a plane passing through the center of a sliding particle, the number
%density $c(\mathbf{r})$ of the solute and the fluid streamlines are shown in 
%Fig.~\ref{fig:reflecting_chi0_0.4}(b).
%Two typical trajectories starting within the basin of
%attraction are illustrated in Figs.~\ref{fig:reflecting_chi0_0.4}(c) and (d). 

The curve containing the saddle point and the attractor is a so-called ``slow manifold'',
characteristic of two-timescale dynamics.\cite{murdock99} In the present case, there is
a separation of timescales between hydrodynamic interaction driven 
slow rotation, and rapid self-phoretic vertical motion. The fast variable $h$
quasi-instantaneously adjusts to the slow variable $\theta$, resulting in the
convergence of trajectories to a quasi-equilibrium curve $U_{z} = 0$ whose 
functional form can be estimated by a simple argument.  
We take the manifold to occur where the vertical component of the \textit{f}ree \textit{s}pace velocity
$U_{fs}\cos(\theta)$ is balanced by additional contributions to $U_{z}$ due to the wall. 
%The chemical gradient created by the wall contributes to $U_{z}$ by modifying the surface slip. 
At leading order, the effect of the wall on $c(\mathbf{r})$ can be modeled as an
image point source with a spatial gradient $\propto1/h^{2}$ (see {\it SI}).  
The leading order hydrodynamic contribution to $U_{z}$ also decays as $1/h^{2}$.\cite{spagnolie12}
%(In contrast, the leading hydrodynamic contribution to $\dot{\theta}$ decays as $1/h^{3}$ \cite{spagnolie12}.)
Thus, we obtain the \textit{q}uasi-\textit{e}quilibrium height $h_{qe}(\theta) \sim (\theta -
\pi/2)^{-1/2}$ near $\theta = \pi/2$. With a fitted prefactor, the predicted slow manifold
is in a very good agreement with the results of the numerical calculations (see the dashed line
in Fig.~\ref{fig:reflecting_chi0_0.4}(a) for $\chi_{0} = 0.4$).

%%%%%%%%%%%%%%%%%%%%%%%%%%%%%%%%%%%%%%%%%%
%\begin{figure}[!tbh]
\begin{figure}[tbh]
\centering
\includegraphics[width=0.95\linewidth]{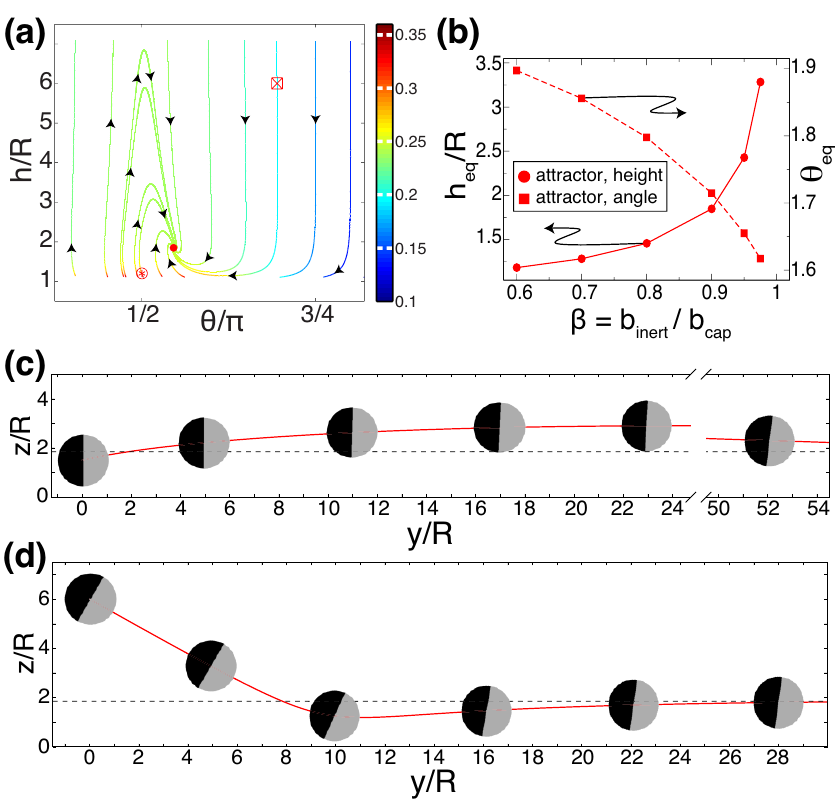}
\caption{\label{fig:chi0_0.0_inert_0.9} (a) Phase plane for half coverage ($\chi_{0} =
0$) and unequal surface mobilities: $\beta = b_{inert}/b_{cap} = 0.9$. There is an
attractor at $h_{eq}/R = 1.85$ and $\theta_{eq} = 98.2^{\circ}$ ({\color{red}$\bullet$}).
Colors indicate $U_{y}/U_{0}$, where $U_{0} \equiv |b_{cap} \kappa/D|$. (b) Variation of
the attractor location with $\beta$ for half coverage. (c) Trajectory with initial
configuration $h_{0}/R = 1.2$ and $\theta_{0} = \pi/2$ (\textcolor{red}{$\circledast$} in (a)). 
% *** change ****
%The particle initially moves away from the wall before rotating its cap away from the wall and
%approaching a steady height and orientation. 
(d) Trajectory with $h_{0}/R = 6$ and
$\theta_{0} = 125^{\circ}$ (\textcolor{red}{$\boxtimes$} in (a)).  This is the same initial
configuration as in Fig. \ref{fig:reflecting_chi0_0.0}(c).}
\end{figure}
%%%%%%%%%%%%%%%%%%%%%%%%%%%%%%%%%%%%%%%%%

Upon increasing $\chi_{0}$ further, we find that $\theta_{eq}$ increases and $h_{eq}$
decreases (see Fig.~\ref{fig:reflecting_chi0_0.4}(e)). For $\chi_{0} > 0.6$, the dynamical
attractor migrates below $h/R = 1.1$.
%, which is outside the region of phase space we consider.  
However, at $\chi_{0} = 0.85$, an attractor emerges at $\theta = \pi$. For this
attractor, $U_{y} = 0$: the particle ``hovers'' in space and acts as a stationary
micropump (see Fig.~\ref{fig:reflecting_chi0_0.9}(b)). 
As shown in Fig.~\ref{fig:reflecting_chi0_0.9}(a), the basin of attraction for
``hovering'' at $\chi_{0} = 0.9$ encompasses nearly half of the phase space.
%; in panel (b) we illustrate the structure of the flow field and of the solute number density corresponding to the hovering state. 
The mechanism of the hovering state is 
understood by balancing the free space velocity with the 
wall-induced contributions to $U_{z}$, dominated by the leading order
solute number density term (the image point source),
%Since the catalyst coverage is very high, we
%expect these contributions to be dominated by the leading order image point source term 
%solute number density term, i.e., the image point source. 
%Restricting our consideration to this term, we obtain 
which gives $h_{eq}/R = [3 (1 - \chi_{0})]^{-1/2}$ (see {\it SI} for details). 
This expression, for most values of $\chi_{0}$, is in good agreement with the numerical
results (see Fig.~\ref{fig:reflecting_chi0_0.9}(e)).

%In the case that the interactions between the solute and the particle are attractive,
%rather than repulsive as considered thus far, it is easy to see that the sole change in
%the equations describing the motion is that the phoretic slip changes sign. Therefore,
%at given coverage $\chi_0$ the phase planes, trajectories, and flow structures can be
%inferred from the ones corresponding to repulsive interactions by simply reversing the
%directions of the arrows. This implies that, for attractive interactions, the sliding and
%hovering attractors turn into repellers, saddle points stay the same, and a particle
%always either ``crashes'' into the wall or moves away from it, in contrast to the complex
%behavior observed in the case of repulsive interactions. 

So far only the case of uniform surface mobility $b<0$ was considered.
%in which the particle rotation is driven solely by the hydrodynamic effect of the wall. 
Here we briefly outline the main effects induced by allowing the ratio $\beta \equiv b_{inert}/b_{cap}$ of the surface
mobilities across the inert and catalytic regions to take values different than one.
Since $\beta \neq 1$ introduces an additional mechanism for
rotating the particle (see {\it SI}), we expect that by adjusting
$\beta$ sliding states can be induced also at the experimentally relevant 
case of half coverage $\chi_0 = 0$.
%(This is the generic case because the Janus particle involves two materials; a detailed
%analysis of this case will be presented elsewhere.) 
%We find that for values $\beta$ slightly different from 1, the structures of the phase planes obtained
%previously, and thus the occurrence of the sliding and hovering states, are preserved (see
%{\it SM}). 
%Secondly, because $\beta \neq 1$ introduces an additional mechanism for
%rotating the particle (see {\it SM}), it is natural to consider whether by adjusting
%$\beta$ sliding states can be induced at the experimentally relevant case of half coverage
%$\chi_0 = 0$. 
For $\beta > 1$ we obtain the reflection only with no dynamical attractor. 
For values $\beta < 1$, however, sliding states occur with $h_{eq},\theta_{eq}$ depending on $\beta$ 
(see Fig.~\ref{fig:chi0_0.0_inert_0.9}). Physically, this occurs because of the stronger
product repulsion from the cap than from the inert side. 
Therefore it becomes possible for the wall-induced chemical gradient to drive a rotation
of the cap \textit{away} from the wall.  

\textcolor{black}{Briefly, we consider the effect of thermal noise, which so far has been neglected. Numerically, we can estimate the ``stiffness'' of a steady state by computing the eigenvalues of the Jacobian at the fixed point $(h^{*}, \theta^{*})$. For the sliding state with $\chi_{0} = 0.4$, we obtain eigenvalues $\lambda_{1} R/U_{0} = -0.011$ and $\lambda_{2} R/U_{0} = -0.145$, reflecting the separation of timescales discussed previously. For the hovering state with $\chi_{0} = 0.9$, we obtain eigenvalues $\lambda_{1} R/U_{0} = -0.014$ and $\lambda_{2} R/U_{0} = -0.059$. Considering a typical catalytic Janus colloid of $R = 2.5 \: \mu m$, half covered by catalyst, and moving, if unconfined, with speed $U_{fs} = 5\:\mu m/s$, one has $U_{0} = 20\:\mu m/s$ (for half coverage, $U_{fs} = U_{0}/4$).\cite{popescu10} Therefore, for both steady states we obtain  $\tau_{1} \approx 10\:s$ as the longest timescale for self-trapping via near-surface swimming. In comparison, for the same colloid the timescale $\tau_{r}$ for reorienting via rotational diffusion is $\tau_{r} = D_{r}^{-1} = 8 \pi \mu R^{3}/k_{B} T$. In water at room temperature this renders as $\tau_{r} \approx 95\:s$. Similarly, the characteristic timescale for translational diffusion (i.e., a perturbation along vertical direction) is $\tau_{t} = R^{2}/D = 6 \pi \mu R^{3}/k_{B} T$, leading to $\tau_{t} \approx 70\:s$. Therefore, for a typical catalytic Janus particle we expect the sliding and hovering states to be robust against thermal noise. For particles with half coverage, we anticipate that both deterministic swimming and rotational diffusion promote the experimentally observed scenario of transient near-surface swimming followed by escape. \cite{volpe11,kreuter13} A study including both effects would be a natural extension of this work.}

Finally, we comment on the case of attractive 
($b>0$) solute-particle interactions. It is easy to see that in this case the sole change in
the equations describing the self-propulsion is that the phoretic slip changes sign.
Therefore,  the phase planes, trajectories, and flow structures can be inferred from
the ones corresponding to $b<0$ by simply reversing the
directions of the arrows. In this case the sliding and hovering attractors will turn 
into repellers, saddle points stay the same, and a particle always either ``crashes'' into
the wall or moves away from it, in contrast to the complex
behavior observed in the case of repulsive interactions. 

To conclude, a catalytically active Janus particle moving near a wall reveals a very rich 
behavior, including reflection, steady sliding, and hovering. \textcolor{black}{Although we have focused on self-diffusiophoresis, we expect, following the
line of reasoning and argumentation presented in Ref. \citenum{golestanian07}, that our results could be relevant for more complex mechanisms of
propulsion, like self-thermophoresis\cite{jiang10,yang14} or self-electrophoresis.\cite{paxton06b,takagi14}} The sliding states could
provide a starting point to establish a stable and predictable motion of swimmers in
microdevices. \textcolor{black}{The sliding mechanism outlined
here could account for the experimental observations reported in Ref.
\citenum{brown14}, namely an accumulation of catalytically active particles near both
the upper and the lower surfaces in capillaries.} Hovering particles create recirculating regions of flow, and could be used
to mix fluid or to trap other particles. Our findings highlight the significant role
played by the wall-induced chemical gradients: due to coupling back to the particle motion
via changing the phoretic slip on the particle surface (neglected in previous studies of
active particles near interfaces), they induce distinct types of motion of the particle.
Finally, we have shown how qualitative and quantitative changes in the behavior can be
achieved in a controlled way by adequately tuning experimentally accessible design
parameters of Janus particles, such as the extent of catalytic coverage $\chi_{0}$ and the
spatial variation of the surface mobility $b$. \textcolor{black}{A detailed study, which considers more general boundary conditions at the wall (such as phoretic slip, or a ``porous'' wall), as well as general values of the parameters $\beta$ and $\chi_{0}$, is currently in progess.}

The authors wish to thank C. Pozrikidis for making freely available the \texttt{BEMLIB} 
library, which was used for the present numerical computations.\cite{pozrikidis02} W.E.U,
M.T., and M.N.P. acknowledge financial support from the DFG, grant No. TA 959/1-1.

%\begin{figure}[h]
%\centering
%  \includegraphics[height=3cm]{example}
%  \caption{An example figure caption \label{fgr:example}}
%\end{figure}

% an example of a two-column figure
%\begin{figure*}
  %\centering
  %\includegraphics[height=3cm]{example.jpg}
  %\caption{An example figure caption, an image from the \textit{Physical Chemistry Chemical Physics} cover gallery.}
  %\label{fgr:example}
%\end{figure*}

%The conclusions section should come at the end of article. For the reference section, the style file rsc.bst can be used to generate the correct reference style.%\footnote[4]{Footnotes should appear here. These might include comments relevant to but not central to the matter under discussion, limited experimental and spectral data, and crystallographic data.}

 %For footnotes in the main text of the article please number the footnotes to avoid duplicate symbols. e.g.  \footnote[num]{your text} the corresponding author \ast counts as footnote 1, ESI as footnote 2, e.g. if there is no ESI, please start at [num]=[2], if ESI is cited in the title please start at [num]=[3] etc. Please also cite the ESI within the main body of the text using \dag.


%\balance

\footnotesize{
\bibliography{rsc_revised} %your .bib file

\providecommand{\noopsort}[1]{}\providecommand{\singleletter}[1]{#1}%
\providecommand*{\mcitethebibliography}{\thebibliography}
\csname @ifundefined\endcsname{endmcitethebibliography}
{\let\endmcitethebibliography\endthebibliography}{}
\begin{mcitethebibliography}{32}
\providecommand*{\natexlab}[1]{#1}
\providecommand*{\mciteSetBstSublistMode}[1]{}
\providecommand*{\mciteSetBstMaxWidthForm}[2]{}
\providecommand*{\mciteBstWouldAddEndPuncttrue}
  {\def\EndOfBibitem{\unskip.}}
\providecommand*{\mciteBstWouldAddEndPunctfalse}
  {\let\EndOfBibitem\relax}
\providecommand*{\mciteSetBstMidEndSepPunct}[3]{}
\providecommand*{\mciteSetBstSublistLabelBeginEnd}[3]{}
\providecommand*{\EndOfBibitem}{}
\mciteSetBstSublistMode{f}
\mciteSetBstMaxWidthForm{subitem}
{(\emph{\alph{mcitesubitemcount}})}
\mciteSetBstSublistLabelBeginEnd{\mcitemaxwidthsubitemform\space}
{\relax}{\relax}

\bibitem[Patra \emph{et~al.}(2013)Patra, Sengupta, Duan, Zhang, Pavlick, and
  Sen]{patra12}
D.~Patra, S.~Sengupta, W.~Duan, H.~Zhang, R.~Pavlick and A.~Sen,
  \emph{Nanoscale}, 2013, \textbf{5}, 1273--1283\relax
\mciteBstWouldAddEndPuncttrue
\mciteSetBstMidEndSepPunct{\mcitedefaultmidpunct}
{\mcitedefaultendpunct}{\mcitedefaultseppunct}\relax
\EndOfBibitem
\bibitem[Paxton \emph{et~al.}(2004)Paxton, Kistler, Olmeda, Sen, St.~Angelo,
  Cao, Mallouk, Lammert, and Crespi]{paxton04}
W.~F. Paxton, K.~C. Kistler, C.~C. Olmeda, A.~Sen, S.~K. St.~Angelo, Y.~Y. Cao,
  T.~E. Mallouk, P.~E. Lammert and V.~H. Crespi, \emph{J. Am. Chem. Soc.},
  2004, \textbf{126}, 13424--13431\relax
\mciteBstWouldAddEndPuncttrue
\mciteSetBstMidEndSepPunct{\mcitedefaultmidpunct}
{\mcitedefaultendpunct}{\mcitedefaultseppunct}\relax
\EndOfBibitem
\bibitem[Paxton \emph{et~al.}(2006)Paxton, Baker, Kline, Wang, Mallouk, and
  Sen]{paxton06b}
W.~Paxton, P.~Baker, T.~Kline, Y.~Wang, T.~Mallouk and A.~Sen, \emph{Angew.
  Chem. Int. Ed.}, 2006, \textbf{128}, 14881--14888\relax
\mciteBstWouldAddEndPuncttrue
\mciteSetBstMidEndSepPunct{\mcitedefaultmidpunct}
{\mcitedefaultendpunct}{\mcitedefaultseppunct}\relax
\EndOfBibitem
\bibitem[Brown and Poon(2014)]{brown14}
A.~Brown and W.~Poon, \emph{Soft Matter}, 2014, \textbf{10}, 4016--4027\relax
\mciteBstWouldAddEndPuncttrue
\mciteSetBstMidEndSepPunct{\mcitedefaultmidpunct}
{\mcitedefaultendpunct}{\mcitedefaultseppunct}\relax
\EndOfBibitem
\bibitem[Ebbens \emph{et~al.}(2014)Ebbens, Gregory, Dunderdale, Howse, Ibrahim,
  Liverpool, and Golestanian]{ebbens14}
S.~Ebbens, D.~A. Gregory, G.~Dunderdale, J.~R. Howse, Y.~Ibrahim, T.~B.
  Liverpool and R.~Golestanian, \emph{EPL}, 2014, \textbf{106}, 58003\relax
\mciteBstWouldAddEndPuncttrue
\mciteSetBstMidEndSepPunct{\mcitedefaultmidpunct}
{\mcitedefaultendpunct}{\mcitedefaultseppunct}\relax
\EndOfBibitem
\bibitem[Gao \emph{et~al.}(2012)Gao, Pei, and Wang]{gao12}
W.~Gao, A.~Pei and J.~Wang, \emph{ACS Nano}, 2012, \textbf{6}, 8432--8438\relax
\mciteBstWouldAddEndPuncttrue
\mciteSetBstMidEndSepPunct{\mcitedefaultmidpunct}
{\mcitedefaultendpunct}{\mcitedefaultseppunct}\relax
\EndOfBibitem
\bibitem[Golestanian \emph{et~al.}(2005)Golestanian, Liverpool, and
  Ajdari]{golestanian05}
R.~Golestanian, T.~B. Liverpool and A.~Ajdari, \emph{Phys. Rev. Lett.}, 2005,
  \textbf{94}, 220801\relax
\mciteBstWouldAddEndPuncttrue
\mciteSetBstMidEndSepPunct{\mcitedefaultmidpunct}
{\mcitedefaultendpunct}{\mcitedefaultseppunct}\relax
\EndOfBibitem
\bibitem[Golestanian \emph{et~al.}(2007)Golestanian, Liverpool, and
  Ajdari]{golestanian07}
R.~Golestanian, T.~B. Liverpool and A.~Ajdari, \emph{New J. Phys.}, 2007,
  \textbf{9}, 126\relax
\mciteBstWouldAddEndPuncttrue
\mciteSetBstMidEndSepPunct{\mcitedefaultmidpunct}
{\mcitedefaultendpunct}{\mcitedefaultseppunct}\relax
\EndOfBibitem
\bibitem[Howse \emph{et~al.}(2007)Howse, Jones, A.J.Ryan, Gough, Vafabakhsh,
  and Golestanian]{howse07}
J.~Howse, R.~Jones, A.J.Ryan, T.~Gough, R.~Vafabakhsh and R.~Golestanian,
  \emph{Phys. Rev. Lett.}, 2007, \textbf{99}, 048102\relax
\mciteBstWouldAddEndPuncttrue
\mciteSetBstMidEndSepPunct{\mcitedefaultmidpunct}
{\mcitedefaultendpunct}{\mcitedefaultseppunct}\relax
\EndOfBibitem
\bibitem[Baraban \emph{et~al.}(2012)Baraban, Tasinkevych, Popescu, Sanchez,
  Dietrich, and Schmidt]{baraban12}
L.~Baraban, M.~Tasinkevych, M.~N. Popescu, S.~Sanchez, S.~Dietrich and O.~G.
  Schmidt, \emph{Soft Matter}, 2012, \textbf{8}, 48--52\relax
\mciteBstWouldAddEndPuncttrue
\mciteSetBstMidEndSepPunct{\mcitedefaultmidpunct}
{\mcitedefaultendpunct}{\mcitedefaultseppunct}\relax
\EndOfBibitem
\bibitem[Ebbens and Howse(2010)]{ebbens10}
S.~J. Ebbens and J.~R. Howse, \emph{Soft Matter}, 2010, \textbf{6},
  726--738\relax
\mciteBstWouldAddEndPuncttrue
\mciteSetBstMidEndSepPunct{\mcitedefaultmidpunct}
{\mcitedefaultendpunct}{\mcitedefaultseppunct}\relax
\EndOfBibitem
\bibitem[Poon(2013)]{poon13}
W.~C.~K. Poon, Proceedings of the International School of Physics ``Enrico
  Fermi'', Course CLXXXIV ``Physics of Complex Colloids'', Amsterdam, 2013, p.
  317\relax
\mciteBstWouldAddEndPuncttrue
\mciteSetBstMidEndSepPunct{\mcitedefaultmidpunct}
{\mcitedefaultendpunct}{\mcitedefaultseppunct}\relax
\EndOfBibitem
\bibitem[Wang \emph{et~al.}(2013)Wang, Duan, Ahmed, Mallouk, and Sen]{wang13}
W.~Wang, W.~Duan, S.~Ahmed, T.~E. Mallouk and A.~Sen, \emph{Nano Today}, 2013,
  \textbf{8}, 531--554\relax
\mciteBstWouldAddEndPuncttrue
\mciteSetBstMidEndSepPunct{\mcitedefaultmidpunct}
{\mcitedefaultendpunct}{\mcitedefaultseppunct}\relax
\EndOfBibitem
\bibitem[Gao \emph{et~al.}(2013)Gao, Feng, Pei, Gu, Li, and Wang]{gao13}
W.~Gao, X.~Feng, A.~Pei, Y.~Gu, J.~Li and J.~Wang, \emph{Nanoscale}, 2013,
  \textbf{5}, 4696--4700\relax
\mciteBstWouldAddEndPuncttrue
\mciteSetBstMidEndSepPunct{\mcitedefaultmidpunct}
{\mcitedefaultendpunct}{\mcitedefaultseppunct}\relax
\EndOfBibitem
\bibitem[Volpe \emph{et~al.}(2011)Volpe, Buttinoni, Vogt, K\"{u}mmerer, and
  Bechinger]{volpe11}
G.~Volpe, I.~Buttinoni, D.~Vogt, H.-J. K\"{u}mmerer and C.~Bechinger,
  \emph{Soft Matter}, 2011, \textbf{7}, 8810--8815\relax
\mciteBstWouldAddEndPuncttrue
\mciteSetBstMidEndSepPunct{\mcitedefaultmidpunct}
{\mcitedefaultendpunct}{\mcitedefaultseppunct}\relax
\EndOfBibitem
\bibitem[Kreuter \emph{et~al.}(2013)Kreuter, Siems, Nielaba, Leiderer, and
  Erbe]{kreuter13}
C.~Kreuter, U.~Siems, P.~Nielaba, P.~Leiderer and A.~Erbe, \emph{Eur. Phys. J.
  Special Topics}, 2013, \textbf{222}, 2923--2939\relax
\mciteBstWouldAddEndPuncttrue
\mciteSetBstMidEndSepPunct{\mcitedefaultmidpunct}
{\mcitedefaultendpunct}{\mcitedefaultseppunct}\relax
\EndOfBibitem
\bibitem[Takagi \emph{et~al.}(2014)Takagi, Palacci, Braunschweig, Shelley, and
  Zhang]{takagi14}
D.~Takagi, J.~Palacci, A.~B. Braunschweig, M.~J. Shelley and J.~Zhang,
  \emph{Soft Matter}, 2014, \textbf{10}, 1784--1789\relax
\mciteBstWouldAddEndPuncttrue
\mciteSetBstMidEndSepPunct{\mcitedefaultmidpunct}
{\mcitedefaultendpunct}{\mcitedefaultseppunct}\relax
\EndOfBibitem
\bibitem[Crowdy(2013)]{crowdy13}
D.~G. Crowdy, \emph{J. Fluid Mech.}, 2013, \textbf{735}, 473--498\relax
\mciteBstWouldAddEndPuncttrue
\mciteSetBstMidEndSepPunct{\mcitedefaultmidpunct}
{\mcitedefaultendpunct}{\mcitedefaultseppunct}\relax
\EndOfBibitem
\bibitem[Popescu \emph{et~al.}(2009)Popescu, Dietrich, and Oshanin]{popescu09}
M.~N. Popescu, S.~Dietrich and G.~Oshanin, \emph{J. Chem. Phys.}, 2009,
  \textbf{130}, 194702\relax
\mciteBstWouldAddEndPuncttrue
\mciteSetBstMidEndSepPunct{\mcitedefaultmidpunct}
{\mcitedefaultendpunct}{\mcitedefaultseppunct}\relax
\EndOfBibitem
\bibitem[Tao and Kapral(2010)]{tao10}
Y.-G. Tao and R.~Kapral, \emph{Soft Matter}, 2010, \textbf{6}, 756--761\relax
\mciteBstWouldAddEndPuncttrue
\mciteSetBstMidEndSepPunct{\mcitedefaultmidpunct}
{\mcitedefaultendpunct}{\mcitedefaultseppunct}\relax
\EndOfBibitem
\bibitem[Lauga \emph{et~al.}(2006)Lauga, Luzio, Whitesides, and Stone]{lauga06}
E.~Lauga, W.~D. Luzio, G.~M. Whitesides and H.~Stone, \emph{Biophys. J.}, 2006,
  \textbf{90}, 400--412\relax
\mciteBstWouldAddEndPuncttrue
\mciteSetBstMidEndSepPunct{\mcitedefaultmidpunct}
{\mcitedefaultendpunct}{\mcitedefaultseppunct}\relax
\EndOfBibitem
\bibitem[Berke \emph{et~al.}(2008)Berke, Turner, Berg, and Lauga]{lauga08}
A.~Berke, L.~Turner, H.~Berg and E.~Lauga, \emph{Phys. Rev. Lett.}, 2008,
  \textbf{101}, 038102\relax
\mciteBstWouldAddEndPuncttrue
\mciteSetBstMidEndSepPunct{\mcitedefaultmidpunct}
{\mcitedefaultendpunct}{\mcitedefaultseppunct}\relax
\EndOfBibitem
\bibitem[Spangolie and Lauga(2012)]{spagnolie12}
S.~Spangolie and E.~Lauga, \emph{J. Fluid Mech.}, 2012, \textbf{700},
  105--147\relax
\mciteBstWouldAddEndPuncttrue
\mciteSetBstMidEndSepPunct{\mcitedefaultmidpunct}
{\mcitedefaultendpunct}{\mcitedefaultseppunct}\relax
\EndOfBibitem
\bibitem[Ishimoto and Gaffney(2013)]{ishimoto13}
K.~Ishimoto and E.~A. Gaffney, \emph{Phys. Rev. E}, 2013, \textbf{88},
  062702\relax
\mciteBstWouldAddEndPuncttrue
\mciteSetBstMidEndSepPunct{\mcitedefaultmidpunct}
{\mcitedefaultendpunct}{\mcitedefaultseppunct}\relax
\EndOfBibitem
\bibitem[Zhang \emph{et~al.}(2010)Zhang, Or, and Murray]{zhang10}
S.~Zhang, Y.~Or and R.~M. Murray, Proc. Am. Control Conf., 2010, pp.
  4205--4210\relax
\mciteBstWouldAddEndPuncttrue
\mciteSetBstMidEndSepPunct{\mcitedefaultmidpunct}
{\mcitedefaultendpunct}{\mcitedefaultseppunct}\relax
\EndOfBibitem
\bibitem[Anderson(1989)]{anderson89}
J.~L. Anderson, \emph{Ann. Rev. Fluid Mech.}, 1989, \textbf{21}, 61--99\relax
\mciteBstWouldAddEndPuncttrue
\mciteSetBstMidEndSepPunct{\mcitedefaultmidpunct}
{\mcitedefaultendpunct}{\mcitedefaultseppunct}\relax
\EndOfBibitem
\bibitem[Popescu \emph{et~al.}(2010)Popescu, Dietrich, Tasinkevych, and
  Ralston]{popescu10}
M.~N. Popescu, S.~Dietrich, M.~Tasinkevych and J.~Ralston, \emph{Eur. Phys. J.
  E}, 2010, \textbf{31}, 351--367\relax
\mciteBstWouldAddEndPuncttrue
\mciteSetBstMidEndSepPunct{\mcitedefaultmidpunct}
{\mcitedefaultendpunct}{\mcitedefaultseppunct}\relax
\EndOfBibitem
\bibitem[Derjaguin \emph{et~al.}(1947)Derjaguin, Sidorenkov, Zubashchenkov, and
  Kiseleva]{derjaguin:47}
B.~V. Derjaguin, G.~P. Sidorenkov, E.~A. Zubashchenkov and E.~V. Kiseleva,
  \emph{Kolloidn. Zh.}, 1947, \textbf{9}, 335\relax
\mciteBstWouldAddEndPuncttrue
\mciteSetBstMidEndSepPunct{\mcitedefaultmidpunct}
{\mcitedefaultendpunct}{\mcitedefaultseppunct}\relax
\EndOfBibitem
\bibitem[Murdock(1999)]{murdock99}
J.~A. Murdock, \emph{Perturbations: Theory and Methods}, SIAM, Philadelphia,
  PA, 1999\relax
\mciteBstWouldAddEndPuncttrue
\mciteSetBstMidEndSepPunct{\mcitedefaultmidpunct}
{\mcitedefaultendpunct}{\mcitedefaultseppunct}\relax
\EndOfBibitem
\bibitem[Jiang \emph{et~al.}(2010)Jiang, Yoshinaga, and Sano]{jiang10}
H.-R. Jiang, N.~Yoshinaga and M.~Sano, \emph{Phys. Rev. Lett.}, 2010,
  \textbf{105}, 268302\relax
\mciteBstWouldAddEndPuncttrue
\mciteSetBstMidEndSepPunct{\mcitedefaultmidpunct}
{\mcitedefaultendpunct}{\mcitedefaultseppunct}\relax
\EndOfBibitem
\bibitem[Yang \emph{et~al.}(2014)Yang, Wysocki, and Ripoll]{yang14}
M.~Yang, A.~Wysocki and M.~Ripoll, \emph{Soft Matter}, 2014, \textbf{10},
  6208--6218\relax
\mciteBstWouldAddEndPuncttrue
\mciteSetBstMidEndSepPunct{\mcitedefaultmidpunct}
{\mcitedefaultendpunct}{\mcitedefaultseppunct}\relax
\EndOfBibitem
\bibitem[Pozrikidis(2002)]{pozrikidis02}
C.~Pozrikidis, \emph{A Practical Guide to Boundary Element Methods with the
  Software Library BEMLIB}, CRC Press, Boca Raton, 2002\relax
\mciteBstWouldAddEndPuncttrue
\mciteSetBstMidEndSepPunct{\mcitedefaultmidpunct}
{\mcitedefaultendpunct}{\mcitedefaultseppunct}\relax
\EndOfBibitem
\end{mcitethebibliography}
\bibliographystyle{rsc} }



\section*{Supplementary material}

# Supplemental Information for
# "Self-propulsion of a catalytically active particle near a planar wall: from reflection to sliding and hovering"

William E. Uspal,[1,2] Mihai N. Popescu,[1,2,3] Siegfried Dietrich,[1,2] and Mykola Tasinkevych[1,2]

[1]*Max-Planck-Institut für Intelligente Systeme, Heisenbergstr. 3, D-70569 Stuttgart, Germany*
[2]*IV. Institut für Theoretische Physik, Universität Stuttgart, Pfaffenwaldring 57, D-70569 Stuttgart, Germany*
[3]*Ian Wark Research Institute, University of South Australia, Adelaide, SA 5095, Australia*

(Dated: November 12, 2014)

## I. CALCULATION OF THE TRANSLATIONAL AND THE ROTATIONAL VELOCITIES OF SELF-PROPELLED PARTICLES

Due to effective solute-particle interactions acting within a thin boundary layer surrounding the particle with a thickness $\delta << R, h$, which is comparable with the range of the interactions, the non-uniform distribution of solute around the colloid drives a surface flow of the solution consisting of solvent and solute. Perpendicular to the particle surface, this flow field characteristically varies over the small length $\delta$. Therefore, it can be modeled as an effective phoretic *slip* velocity $\mathbf{v}_s = -b(\mathbf{r})\nabla_{||}c(\mathbf{r})$, where $\nabla_{||} = (\mathbf{1} - \mathbf{nn}) \cdot \nabla$ is the projection of the gradient onto the particle surface [1, 2], with $\mathbf{n}$ denoting the surface normal oriented towards the fluid. The quantity $b(\mathbf{r})$ is a so-called "surface mobility" determined by the effective interaction between the product and the particle surface [1, 2]. Except where indicated, we take $b(\mathbf{r})$ to be uniform over the surface: $b(\mathbf{r}) = b$. The sign of $b$ tells whether the product is effectively repelled or attracted by the particle. Here we analyze in detail the case of repulsive interactions, i.e., $b < 0$ and briefly summarize the changes occurring in the case of attractive interactions.

Under the assumption that a quasi-steady state establishes fast, and for sufficiently small self-propulsion velocities $\mathbf{U}$ and $\mathbf{\Omega}$ (see below), the number density $c(\mathbf{r})$ is governed by the Laplace equation $D\nabla^2 c = 0$, with the boundary conditions $\partial c/\partial n = 0$ on the planar wall, $\partial c/\partial n = 0$ on the inert surface of the colloid, and $-D(\partial c/\partial n) = \kappa$ on the catalytic cap. $D$ is the diffusion coefficient of the product molecules, i.e., solute, and the constant $\kappa$ is the flux of product molecules per area emanating from the cap. The last boundary condition corresponds to the so-called "constant flux" condition for which the turnover rate of the catalytic reaction is independent of the local concentration of the reactant (as it is, e.g., the case for a Michaelis-Menten reaction mechanism at high reactant concentration or the case if the reaction rate is the limiting step.) The velocity field $\mathbf{u}(\mathbf{r})$ of the solution is governed by the Stokes equation $-\nabla P + \eta \nabla^2 \mathbf{u} = 0$, where $\eta$ is the viscosity of the solution and $P(\mathbf{r})$ is the pressure field, along with the incompressibility condition $\nabla \cdot \mathbf{u} = 0$. For the velocity field, there is a no-slip boundary condition $\mathbf{u}(\mathbf{r}) = 0$ at the planar wall. At the surface of the colloid, the boundary condition for $\mathbf{u}$ is $\mathbf{u}(\mathbf{r}) = \mathbf{U} + \mathbf{\Omega} \times (\mathbf{r} - \mathbf{r}_0) + \mathbf{v}_s$, where $\mathbf{r}_0$ is the position of the center of the sphere and $\mathbf{r}$ is the position of a point on the surface of the particle, while far from the colloid the fluid is at rest. Finally, we require that the colloid, which moves by diffusiophoresis, is force and torque free [1–3], which fixes $\mathbf{U}$ and $\mathbf{\Omega}$.

This formulation of the problem implicitly assumes several approximations. The first one is that we neglect any rotational diffusion effects, i.e., we assume that the rotational diffusion time is much larger than the timescales over which we study the motion. In using the Laplace equation for the number density, we have neglected convection and we have taken the number density to be quasi-static. This approximation is valid if the Peclet number $Pe \equiv U_0 R/D$ is small, where $U_0$ is a characteristic particle velocity. In using the Stokes equation, we have neglected inertia of the solution, which is valid for small Reynolds number $Re \equiv \rho U_0 R/\eta$, where $\rho$ is the mass density of the solution. These characteristic numbers have been estimated as $Re \approx 10^{-5}$ and $Pe \approx 10^{-2}$ for a 10 $\mu m$ particle moving at 1 $\mu m/s$ and catalyzing the decomposition of $H_2O_2$ into $H_2O$ and $O_2$ [3]. We also neglect any phoretic slip induced at the wall by the product. Finally, we assumed that the interfacial length scale $\delta$ is small compared with the particle-surface gap distance; however, we note that for ionic interactions in aqueous solution the range of interaction can be up to hundreds of nm and therefore in this latter case the region of the phase plane where the assumptions of our model are expected to hold would be somewhat smaller, i.e., $h/R \simeq 1.4$ rather than the value 1.1 considered in the main text.

We express all lengths in units of the particle radius $R$; the reaction product number density $c(\mathbf{r})$ in units of $c_0 = \kappa R/D$; the translational velocity in units of $U_0 = |b|\kappa/D$; and the angular velocity in units of $\Omega_0 = U_0 R$.

In order to calculate numerically $\mathbf{U}$ and $\mathbf{\Omega}$ for a colloid configuration (i.e., position $h$ and orientation $\theta$), we take advantage of the fact that the number density of the solute and the velocity field $\mathbf{u}(\mathbf{r})$ of the solution are coupled

solely through the phoretic slip boundary condition for $\mathbf{u}$ at the surface of the colloid. Accordingly, we employ the following steps.

**(i)** First, the number density $c(\mathbf{r})$ is calculated by solving numerically the diffusion equation, subject to the corresponding boundary conditions, using the boundary element method (BEM) implemented with the BEMLIB library [4].

**(ii)** Using the calculated $c(\mathbf{r})$, the phoretic slip $\mathbf{v}_s = -b\nabla_{||}c(\mathbf{r})$ is determined by numerical differentiation of $c(\mathbf{r})$.

**(iii)** $\mathbf{U}$ and $\mathbf{\Omega}$ enter into the boundary conditions. Thus they can be fixed, by requiring that the particle is force- and torque-free, only after the solution of the Stokes equations has been determined. Therefore it is difficult to calculate them numerically by directly solving the initial problem. We shall circumvent this difficulty and determine them by employing the Lorentz reciprocal theorem, which relates the solutions of two distinct Stokes flow problems (called "primed" and "unprimed") which share the same geometry but have different boundary conditions [5]. We proceed as follows:

**(iii)(a)** According to the Lorentz reciprocal theorem, the fluid stresses $(\boldsymbol{\sigma}, \boldsymbol{\sigma}')$ and the velocity fields $(\mathbf{u}, \mathbf{u}')$ of the two problems are related by a surface integral over the fluid domain boundaries:

$$\int \mathbf{u} \cdot \boldsymbol{\sigma}' \cdot \mathbf{n}\, dS = \int \mathbf{u}' \cdot \boldsymbol{\sigma} \cdot \mathbf{n}\, dS\,. \tag{1}$$

The "unprimed" problem is the self-propelled particle under study, for which the boundary conditions are: $\mathbf{u} = 0$ on the planar wall, $\mathbf{u} = \mathbf{U} + \mathbf{\Omega} \times (\mathbf{r} - \mathbf{r}_0) + \mathbf{v}_s$ on the particle surface, and $\mathbf{u} = 0$ at infinity. Note that $\mathbf{v}_s$ has been determined in step **(ii)**, and thus it is a known quantity.

Since there are six unknowns (the six components of the vectors $\mathbf{U}$ and $\mathbf{\Omega}$), six dual ("primed") problems are needed. The choice of these problems is made such that, for the given geometry of the boundaries, they obey boundary conditions for which either the solution is known or can be computed numerically in a straightforward manner.

**(iii)(b)** The six primed problems (indexed by $j = 1, \ldots, 6$) which we consider are the translation with velocity $U_0$ or the rotation with angular velocity $\Omega_0 = U_0 R$ of a spherical particle of radius $R$ through the fluid along each of the three directions $\hat{x}$, $\hat{y}$, and $\hat{z}$ at height $h$ above the planar wall. Choosing $j = 1, 2, 3$ to denote the cases of translation along the directions $\hat{x}$, $\hat{y}$, and $\hat{z}$, respectively, and $j = 4, 5, 6$ the corresponding cases of rotations along $\hat{x}$, $\hat{y}$, and $\hat{z}$ for a sphere moving through the fluid at constant height $h$ above a planar wall, the problem $j$ thus correspond to translational and rotational velocities $(\mathbf{U}'_j, \mathbf{\Omega}'_j)$ as follows:

$$\begin{aligned}
(\mathbf{U}'_{j=1}, \mathbf{\Omega}'_{j=1}) &= (U_0 \mathbf{e}_x, 0), \\
(\mathbf{U}'_{j=2}, \mathbf{\Omega}'_{j=2}) &= (U_0 \mathbf{e}_y, 0), \\
(\mathbf{U}'_{j=3}, \mathbf{\Omega}'_{j=3}) &= (U_0 \mathbf{e}_z, 0), \\
(\mathbf{U}'_{j=4}, \mathbf{\Omega}'_{j=4}) &= (0, \Omega_0 \mathbf{e}_x), \\
(\mathbf{U}'_{j=5}, \mathbf{\Omega}'_{j=5}) &= (0, \Omega_0 \mathbf{e}_y), \\
(\mathbf{U}'_{j=6}, \mathbf{\Omega}'_{j=6}) &= (0, \Omega_0 \mathbf{e}_z),
\end{aligned} \tag{2}$$

where $\mathbf{e}_{x,y,z}$ denote the unit vectors of the $x$, $y$, and $z$ directions, respectively. For each of the cases $j = 1, \ldots, 6$ we impose that the motion is subject to no-slip boundary conditions at the planar wall, i.e., $\mathbf{u}'_j = 0$ at the planar wall, and at the surface of the particle, i.e., $\mathbf{u}'_j = \mathbf{U}'_j + \mathbf{\Omega}'_j \times (\mathbf{r} - \mathbf{r}_0)$ at the particle surface, and that the fluid is quiescent far away from the particle, i.e., $\mathbf{u}'_j = 0$ at infinity. Each of the problems $j = 1, \ldots, 6$ as defined above can be straightforwardly solved numerically using the BEM [4] and therefore the corresponding fluid stresses $\boldsymbol{\sigma}'_j$ and velocity fields $\mathbf{u}'_j$, are known quantities.

**(iii)(c)** We apply the Lorentz theorem (Eq. (1)) to each of the six pairs obtained by combining the unprimed problem with the problem $j$. Far away from the particle and at the planar wall both $\mathbf{u}$ and $\mathbf{u}'_j$ vanish so that concerning the integral over the whole boundary of the fluid domain only the part over the surface of the particle contributes. This leads to the following set of equations:

$$\int_{|\mathbf{r}|=R} \mathbf{u} \cdot \boldsymbol{\sigma}'_j \cdot \mathbf{n}\, dS = \int_{|\mathbf{r}|=R} \mathbf{u}'_j \cdot \boldsymbol{\sigma} \cdot \mathbf{n}\, dS,\; j = 1, \ldots, 6\,. \tag{3}$$

Upon inserting the boundary conditions $\mathbf{u}'_j$ into Eq. (3), on its rhs one obtains

$$\int_{|\mathbf{r}|=R} [\mathbf{U}'_j + \mathbf{\Omega}'_j \times (\mathbf{r} - \mathbf{r}_0)] \cdot \boldsymbol{\sigma} \cdot \mathbf{n}\, dS.$$

We consider the two terms on the rhs in turn. For the translational term, we have

$$\int_{|\mathbf{r}|=R} \mathbf{U}'_j \cdot \boldsymbol{\sigma} \cdot \mathbf{n}\, dS = \mathbf{U}'_j \cdot \int_{|\mathbf{r}|=R} \boldsymbol{\sigma} \cdot \mathbf{n}\, dS = \mathbf{U}'_j \cdot \mathbf{F}\,, \tag{4}$$

where $\mathbf{F} = \int_{|\mathbf{r}|=R} \boldsymbol{\sigma} \cdot \mathbf{n}\, dS$ is, by definition [5], the force exerted by the fluid on the self-propelled particle (plus its thin boundary layer of thickness $\delta$). Since this is the only force acting on the particle, and since the self-propelled particle is force-free ($\mathbf{F} = 0$), this term thus vanishes. For the rotational term, we have

$$\int_{|\mathbf{r}|=R} \mathbf{\Omega}'_j \times (\mathbf{r} - \mathbf{r}_0) \cdot \boldsymbol{\sigma} \cdot \mathbf{n}\, dS = \mathbf{\Omega}'_j \cdot \int_{|\mathbf{r}|=R} (\mathbf{r} - \mathbf{r}_0) \times \boldsymbol{\sigma} \cdot \mathbf{n}\, dS = \mathbf{\Omega}'_j \cdot \boldsymbol{\tau}. \tag{5}$$

Using the vector identity $(\mathbf{a} \times \mathbf{b}) \cdot \mathbf{c} = \mathbf{a} \cdot (\mathbf{b} \times \mathbf{c})$ we have rearranged the integrand and identified the last integral with the torque exerted by the fluid on the self-propelled particle (plus its thin boundary layer of thickness $\delta$) [5]. Since the self-propelled particle is torque-free ($\boldsymbol{\tau} = 0$), the rotational term thus vanishes, too. Therefore, the entire right hand side of Eq. (3) is zero. Upon inserting the boundary conditions for $\mathbf{u}$ into the lhs of Eq. (3), we obtain

$$\int_{|\mathbf{r}|=R} \mathbf{U} \cdot \boldsymbol{\sigma}'_j \cdot \mathbf{n}\, dS + \int_{|\mathbf{r}|=R} \mathbf{\Omega} \times (\mathbf{r} - \mathbf{r}_0) \cdot \boldsymbol{\sigma}'_j \cdot \mathbf{n}\, dS = -\int_{|\mathbf{r}|=R} \mathbf{v}_s \cdot \boldsymbol{\sigma}'_j \cdot \mathbf{n}\, dS. \tag{6}$$

Due to manipulations similar to the above ones, we obtain

$$\mathbf{U} \cdot \mathbf{F}'_j + \mathbf{\Omega} \cdot \boldsymbol{\tau}'_j = -\int_{|\mathbf{r}|=R} \mathbf{v}_s \cdot \boldsymbol{\sigma}'_j \cdot \mathbf{n}\, dS\,, \quad j = 1, \ldots, 6\,, \tag{7}$$

with $\mathbf{F}'_j = \int_{|\mathbf{r}|=R} \boldsymbol{\sigma}'_j \cdot \mathbf{n}\, dS$ and $\boldsymbol{\tau}'_j = \int_{|\mathbf{r}|=R} (\mathbf{r} - \mathbf{r}_0) \times \boldsymbol{\sigma}'_j \cdot \mathbf{n}\, dS$. $\mathbf{F}'_j$ and $\boldsymbol{\tau}'_j$ are the force and torque, respectively, exerted by the quiescent fluid on the particle in steady-state translation (or rotation) with a no-slip boundary condition on its surface. This state is possible only if the particle is driven; therefore $\mathbf{F}'_j \neq 0, \boldsymbol{\tau}'_j \neq 0$.

Since $\mathbf{v}_s$ (see **(ii)**) and the fluid stresses $\boldsymbol{\sigma}'_j$ are known (from the numerical solution of each dual problem, see **(iii)(b)**), Eq. (7) provides a system of six linear equations for the six unknown translational ($\mathbf{U}$) and rotational ($\mathbf{\Omega}$) velocity components. This completes the calculation which in the end amounts to solving a system of six linear equations.

## II. CONTRIBUTION OF SOLUTE GRADIENTS TO PARTICLE ROTATION

Briefly, we show that for a particle with uniform surface mobility, gradients in solute density cannot directly contribute to the angular velocity of a spherical, self-diffusiophoretic particle. We use the reciprocal theorem for a particle in unconfined fluid. Without loss of generality, we take our dual ("primed") problem to be a sphere rotating around the $\hat{z}$ axis with angular velocity $\Omega'_z$ in quiescent fluid. We wish to find the component of angular velocity $\Omega_z$ of an unconfined particle in an arbitrary solute number density field $c(\mathbf{r})$. The torque on the rotating sphere is $\tau'_z = -8\pi\eta R^3 \Omega'_z$:

$$-8\pi\eta R^3 \Omega'_z \Omega_z = -\int_{|\mathbf{r}|=R} \mathbf{v}_s \cdot \boldsymbol{\sigma}' \cdot \mathbf{n} \, dS. \tag{8}$$

We introduce a spherical coordinate system centered on the particle center. Of the tangential components of the stress tensor, $\sigma'_{\theta r} = 0$, leaving only $\sigma'_{\phi r} = -3\eta \Omega'_z \sin(\theta)$ [6]. Invoking $\mathbf{v}_s = -b(\mathbf{r})\nabla_\parallel c(\mathbf{r})$, we have:

$$8\pi\eta R^3 \Omega'_z \Omega_z = \int_{|\mathbf{r}|=R} b(\mathbf{r}) \left(\frac{1}{R\sin(\theta)} \frac{\partial c}{\partial \phi}\right) (3\eta \Omega'_z \sin(\theta)) R^2 \sin(\theta) \, d\theta d\phi, \tag{9}$$

$$\Omega_z = \frac{3}{8\pi R} \int_{|\mathbf{r}|=R} b(\mathbf{r}) \left(\frac{\partial c}{\partial \phi}\right) \sin(\theta) \, d\theta d\phi. \tag{10}$$

If the surface mobility $b(\mathbf{r})$ is uniform, $b(\mathbf{r}) = b$, then $\Omega_z = 0$, because $c(\mathbf{r})$ is a single-valued function. Since the choice of rotation axis in the dual problem was arbitary, we have $\boldsymbol{\Omega} = 0$.

These results are not surprising. Smoluchowksi demonstrated that, for an (inert) spherical particle with uniform surface mobility in an unbounded fluid, a linearly varying solute density *externally* maintained cannot drive phoretic rotation of the particle [1]. This result was extended by Keh and Anderson for an arbitrarily varying external solute density [1].

In order to extend this rigorous analysis to the case of a planar wall detailed, careful consideration of Eq. (7) would be required because in the vicinity of a wall translation and rotation of a particle are hydrodynamically coupled. Nevertheless, one can acquire insight about the strength of the wall effects by examining the influence of the wall on the terms in Eq. (8), only.

We have shown that, in free space, the integral in Eq. (10) is zero for any distribution of solute, provided that the surface mobility of the particle is uniform. Hence, for a particle to rotate in the vicinity of a wall, the wall must modify the stress tensor $\boldsymbol{\sigma}'$. For a *rotating* sphere, the leading order, wall-induced contributions to $\boldsymbol{\sigma}'$ stem from an image rotlet (point torque) and image stresslet [7]. The stress from these singularities decays as $1/r^3$. Hence, we anticipate that in the vicinity of a wall a contribution of the order $1/h^3$ to the angular velocity of the particle may occur. This is consistent with the well-known fact that for a self-propelled sphere near a wall the sphere can rotate with an angular velocity proportional to $1/h^3$ via the hydrodynamic interaction with the wall, i.e., the confinement of the disturbance flow created by the particle [8]. The reciprocal theorem (Eq. (7)) thus inherently captures the hydrodynamic interaction with the wall.

In order to determine whether wall-induced chemical gradients can rotate a particle, we must also consider the effect of the wall on $\mathbf{v}_s$ through $c(\mathbf{r})$. At the simplest level, the effect of a wall on the solute density field can be modeled as an image point source of solute. Gradients from a point source decay as $1/r^2$, thus the image source changes the slip velocity by a term of the order $1/h^2$. Therefore, combining these two estimates, we anticipate that wall-induced chemical gradients could contribute to a rotation of the particle at order $1/h^5$, *but only through a coupling with the reflected flows*.

For $h/R \geq 2$, one has $(R/h)^5 \leq 0.032$, which can be regarded as negligible. Thus such contributions of the order $(R/h)^5$ would be relevant only in the region very near the wall. Since we consider particle-wall separations as small as $h/R = 1.1$, for which $(R/h)^5 \approx 0.62$, further analysis is needed. Accordingly, in the next section, we consider a numerical decomposition of chemical and hydrodynamic contributions to the rotation of the particle.

## III. NUMERICAL IDENTIFICATION OF THE WALL CONTRIBUTIONS TO THE PARTICLE MOTION

Equation (7) includes the effects of the wall in two distinct ways. The stress tensors $\boldsymbol{\sigma}'_j$ and the associated forces $\mathbf{F}'_j$ and torques $\boldsymbol{\tau}'_j$ include the *hydrodynamic* effect of the wall (i.e., the consequences of imposing the no-slip boundary condition $\mathbf{u} = 0$ at the wall), while the slip velocity $\mathbf{v}_s$ at the particle surface includes the *chemical* effect of the wall (i.e., the consequences of imposing the no-flux condition $\partial c/\partial n = 0$ at the wall). Via substitution of suitable alternatives (see below) for these quantities, we can identify the roles of the two effects in determining the particle motion. In order to identify the chemical effect of the wall, we can use as the dual problem a particle translating or rotating in free space, and substitute the *free space* stress tensor $\boldsymbol{\sigma}'^{fs}_j$ and the associated forces and torques into Eq. (7). In order to identify the hydrodynamic effect of the wall, one can calculate the slip velocity $\mathbf{v}^{fs}_s$ for the active particle in free space, and substitute this quantity into the rhs of Eq. (7).

Now we consider the wall-induced chemical and hydrodynamic contributions to $\dot\theta = -\Omega_x$ for half coverage ($\chi_0 = 0$) (note that the minus sign is due to the definition of $\theta$, see Fig. 1(a) in the main text). Strikingly, the chemical contribution of the wall to rotation is orders of magnitude smaller than the hydrodynamic contribution (Fig. 1). This is expected from the theoretical arguments advanced in the previous section. Therefore the rotation appearing in the trajectories in Figs. (1)-(3) in the main text are solely due to the hydrodynamic interaction of the particle with the planar wall. In fact, because we expect the chemical contribution to be identically zero, Fig. 1(c) indicates the accuracy of our numerical scheme. For $h/R$ close to the lower bound $h/R = 1.1$, the accuracy of the numerically predicted quantity $\dot\theta/\Omega_0$ is of the order of $10^{-5}$. For $h/R = 1.5$, the accuracy increases to $10^{-7}$.

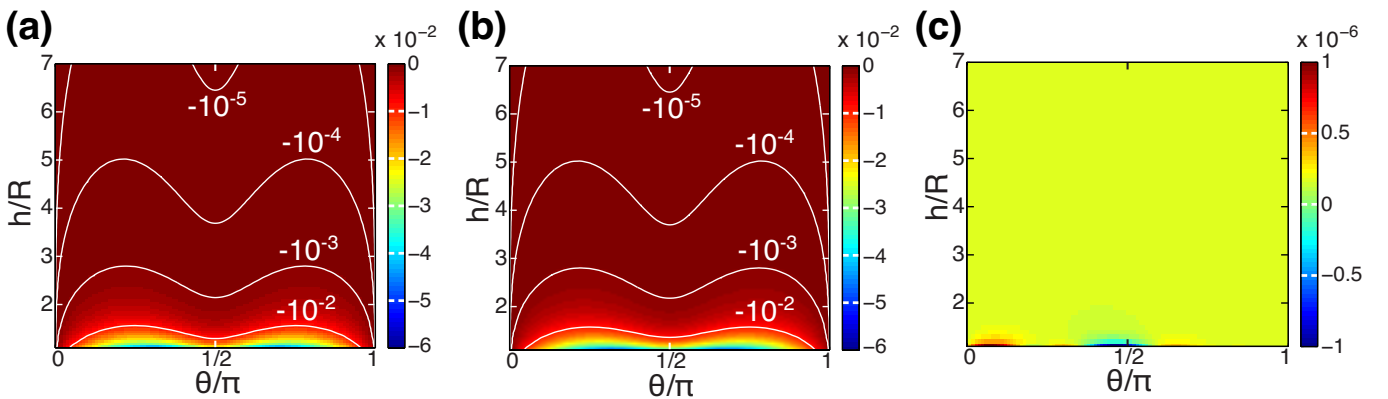

FIG. 1: (a) $\dot\theta$ (in units of $\Omega_0$) as function of the particle configuration ($h/R,\theta$) for half coverage ($\chi_0 = 0$) (see the color code and the white lines of constant $\dot\theta$). This plot includes both the chemical and hydrodynamic effects of the wall: since the wall is impenetrable to the product, it creates a gradient in number density; and since there is a no-slip condition at the wall, it reflects the disturbance flows created by the particle. (b) Plot of $\dot\theta/\Omega_0$ which is obtained by using the free space slip velocity $\mathbf{v}^{fs}_s$ in the context of the reciprocal theorem, i.e., neglecting the wall-induced effect on the product number density, but including the effect of the wall on the hydrodynamics. The plot, which is symmetric around $\theta/\pi = 1/2$, is virtually indistinguishable from (a). (c) $\dot\theta/\Omega_0$ obtained by using the free space stress tensor $\boldsymbol{\sigma}'^{fs}_j$ in the reciprocal theorem, i.e., by neglecting the hydrodynamic effect of the wall, but including its effect on the number density. The angular velocity $\dot\theta/\Omega_0$ is orders of magnitude lower than in (a) and (b). Thus we conclude that wall-induced number density gradients do not drive the rotation of the particle.

Furthermore, we note that Fig. 1(a) and Fig. 1(b) are symmetric around $\theta = \pi/2$. Nevertheless, the phase plane trajectories in Fig. 1(b) in the main text are clearly asymmetric. This asymmetry is due to the chemical effect of the wall on $U_z$. In Fig. 2, we show the phase plane for half coverage calculated with $\mathbf{v}^{fs}_s$, i.e., with the effect of the wall on the number density of solute neglected. These trajectories more closely resemble those obtained from theoretical models in which the surface slip is specified *a priori* [8, 9]. In particular, a particle initially facing the wall with an orientation $\theta_0 = \pi/2 + \Delta\theta$ at a height $h_0$ approaches the wall, reaches a turning point, and leaves the wall such that it has the orientation $\theta = \pi/2 - \Delta\theta$ when it reaches the height $h_0$ again. In comparison with Fig. 1(b) in the main text, the trajectories still fall into three classes of behavior: direct escape, initial approach to the wall and subsequent reflection, and "crashing". However, the reflected trajectories do not evolve into a unique asymptotic orientation as $t \to \infty$, unlike those in the region $\pi/2 < \theta < 0.7\pi$ in Fig. 1(b) in the main text, which end up with $\theta \simeq \pi/2$.

However, it should be noted that our method to decompose the angular velocity function $\dot\theta(h,\theta)$ into hydrodynamic and chemical contributions is not exact. That is, for any particle surface chemistry, adding plots like those in Figs.

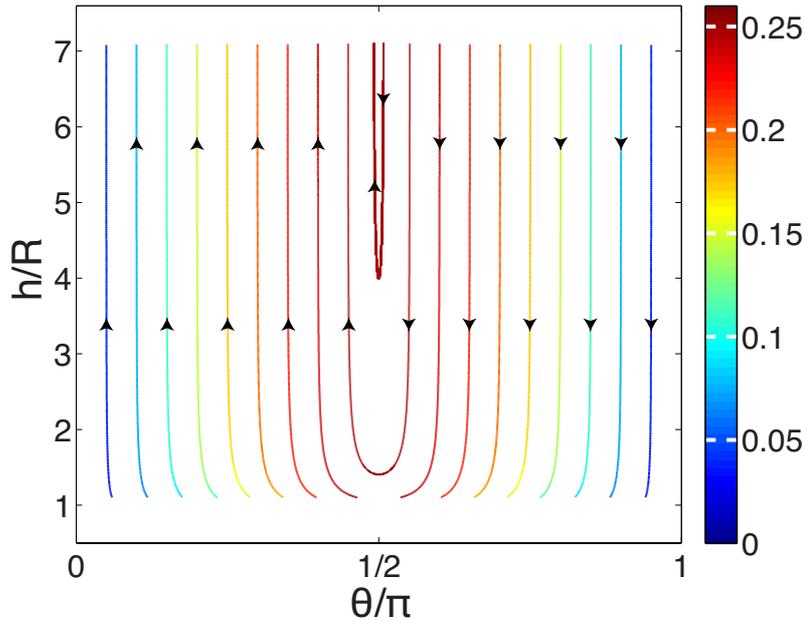

FIG. 2: Phase plane obtained for half coverage ($\chi_0 = 0$) and using the free space slip velocity $\mathbf{v}_s^{fs}$ in the reciprocal theorem, i.e., by neglecting the effect of the wall on the product number density. The trajectories of the particle are symmetric about $\theta = \pi/2$; there is no specific orientation the scattered particle evolves to asymptotically, unlike $\theta \simeq \pi/2$ in Fig. 1(b) in the main text. Color indicates $U_y/U_0$.

1(b) and (c) do not exactly reproduce a plot such as Fig. 1(a), even in the absence of numerical errors. One reason is that the wall-induced chemical gradient modifies the surface slip velocity, leading to a coupling of hydrodynamic and chemical effects at higher orders in $1/h$. We can estimate how the error associated with simply adding (b) and (c) scales with the particle height as follows.

As discussed in the previous section, the presence of a boundary which is impenetrable to the solute modifies the surface slip velocity with a contribution which decays as $1/h^2$. The surface slip distribution determines the strength of the hydrodynamic singularities which characterize how an active particle disturbs the surrounding fluid. For a force- and torque-free particle, the leading order hydrodynamic singularity is a force dipole. If the force dipole strength $\alpha$ for an unconfined (free space) particle is $\alpha_{fs}$, then the strength in the presence of a impenetrable wall carries a correction $\sim 1/h^2$: $\alpha \approx \alpha_{fs} + \alpha_2/h^2$. In the presence of a no-slip boundary at the wall, the reflection of the disturbance flow created by a force dipole contributes to the vertical (wall normal) component of the particle velocity with an effect which decays as $1/h^2$ [8]. Therefore, the contribution of the correction $\alpha_2/h^2$ to the *vertical* velocity is expected to decay as $1/h^4$. On the other hand, the correction to the *angular* velocity is expected to decay as $1/h^5$ as discussed in the previous section.

Accordingly, it is legitimate to ask how to recognize and assess the significance of higher order effects. They can be quantified by finding the difference between the complete angular velocity, (Fig. 1(a)), and the sum of the isolated contributions (Fig. 1(b) and Fig. 1(c)). However, since the chemical contribution is negligibly small, as shown in Fig. 1(c), we can directly compare Fig. 1(a) and Fig. 1(b). Except for being very close to the wall and near $\theta = 90°$, the two functions are virtually identical. We conclude that rotation is driven largely by hydrodynamic interactions. A detailed study of higher order effects will be presented elsewhere.

## IV. ANALYTICAL ESTIMATE OF HOVERING HEIGHT

We seek to obtain an expression for the equilibrium height $h_{eq}$ of the hovering state as a function of the catalyst coverage $\chi_0$. Since for this state the particle catalyst coverage is very high, we expect the chemical effect of the wall to dominate the hydrodynamic interaction between the particle and the wall, which we therefore neglect.

The number density of solute created by the particle can be expressed in terms of a multipole expansion. The monopole term in this expansion captures the total flux of product emanating from the particle. Higher order terms are caused by the anisotropic distribution of catalyst on the surface of the particle. In order to impose the no-flux boundary condition at the wall, each term in the multipole expansion is associated with an image singularity located below the wall at $z = -h$, using the coordinate system of Fig. 1(a) in the main text.

We switch to a coordinate system in which the particle center is located at the origin, so that the image singularities are located at $z = -2h$. In the present context we restrict our attention to the image monopole. The contribution of this singularity to the full number density is

$$c_m^*(\mathbf{r}) = \frac{\dot{m}}{4\pi D |\mathbf{r} + 2h\hat{\mathbf{z}}|}, \tag{11}$$

where $\dot{m}$ is the total flux of product from the particle surface, $\hat{\mathbf{z}}$ is the normal of the planar wall, and the asterisk indicates that this is an image singularity. The total flux $\dot{m}$ is related to the coverage parameter by $\dot{m} = 2\pi R^2 \kappa (1+\chi_0)$, so that

$$c_m^*(\mathbf{r}) = \frac{\kappa R^2 (1+\chi_0)}{2D |\mathbf{r} + 2h\hat{\mathbf{z}}|}. \tag{12}$$

Now we seek the surface gradient $\nabla_\parallel c_m^*(\mathbf{r})$. Invoking the rotational symmetry around the $\hat{z}$ axis (i.e., $c_m^*$ does not depend on the azimuthal angle $\phi$), the surface gradient is $\nabla_\parallel c_m^*(\mathbf{r}) = [\nabla c_m^*(\mathbf{r}) \cdot \mathbf{e}_\vartheta] \mathbf{e}_\vartheta$, where $\vartheta$ denotes the polar angle in spherical coordinates and $\mathbf{e}_\vartheta$ is the unit vector corresponding to the $\vartheta$ direction. We obtain

$$\nabla_\parallel c_m^*(\mathbf{r}) = \left( \frac{-\kappa R^2 (1+\chi_0)(\mathbf{r} + 2h\hat{\mathbf{z}}) \cdot \mathbf{e}_\vartheta}{2D |\mathbf{r} + 2h\hat{\mathbf{z}}|^3} \right) \mathbf{e}_\vartheta = \frac{h\kappa R^2 (1+\chi_0)}{D |\mathbf{r} + 2h\hat{\mathbf{z}}|^3} (\sin \vartheta) \, \mathbf{e}_\vartheta \,, \tag{13}$$

where the second equality follows from the fact that $\mathbf{r}$ and $\mathbf{e}_\vartheta$ are orthogonal and $\hat{\mathbf{z}} \cdot \mathbf{e}_\vartheta = -\sin \vartheta$.

In order to evaluate the velocity of the particle along the $\hat{\mathbf{z}}$ direction we apply the reciprocal theorem (Eq. (7)). For analytical tractability, in the dual problem we use the stress tensor $\boldsymbol{\sigma}'$ for a particle moving with velocity $U_z'$ in the $\hat{\mathbf{z}}$ direction in free space. This is an approximation which we expect to be valid far from the wall. In this case, on the lhs of Eq. (7) there is no torque exerted by the fluid on the particle, i.e., $\boldsymbol{\tau}' = 0$, while the force $\mathbf{F}'$ exerted by the fluid on the particle is simply given by the Stokes formula for a sphere, $\mathbf{F}' = -6\pi \eta R U_z' \, \hat{\mathbf{z}}$ [5]. One therefore obtains

$$-6\pi \eta R U_z' U_z^{*,m} = -\int_{|\mathbf{r}|=R} \mathbf{v}_s \cdot \boldsymbol{\sigma}' \cdot \mathbf{n} \, dS, \tag{14}$$

where $U_z^{*,m}$ is the velocity component we wish to obtain (the superscript $(*,m)$ being a reminder that the velocity $U_z$ is calculated by using an approximation of the wall effect on the number density which accounts only for the image monopole). The slip velocity induced by the image monopole is $\mathbf{v}_{s,m}^* = -b\nabla_\parallel c_m^*(\mathbf{r})$ (for a uniform phoretic mobility $b(\mathbf{r}) = b$). Since according to Eq. (13) the slip velocity has only a component in the direction of $\mathbf{e}_\vartheta$ while the normal to the surface of the sphere is along the radial direction, $\hat{\mathbf{n}} = \hat{\mathbf{r}} = \mathbf{r}/r$, only the component $\sigma'_{r\vartheta}$ of the stress tensor $\boldsymbol{\sigma}'$, evaluated at the surface of the sphere, is relevant for the integral on the rhs of Eq. (14). For a sphere translating

with velocity $U'_z$ through an unbounded fluid $\sigma'_{r\vartheta}(R,\vartheta) = 3\eta U'_z \sin\vartheta/(2R)$ [5, 10], and thus we obtain

$$6\pi \eta R U'_z U_z^{*,m} = -\int_0^\pi d\vartheta \left[\frac{bh\kappa R^2(1+\chi_0)}{D|R\hat{\mathbf{r}} + 2h\hat{\mathbf{z}}|^3} \sin\vartheta\right]\left(\frac{3\eta U'_z}{2R} \sin\vartheta\right) 2\pi R^2 \sin\vartheta, \tag{15}$$

which implies

$$\begin{aligned}
U_z^{*,m} &= -\int_0^\pi d\vartheta \frac{bh\kappa R^2(1+\chi_0)}{2D|R\hat{\mathbf{r}} + 2h\hat{\mathbf{z}}|^3}(\sin\vartheta)^3 \\
&= -\frac{b\kappa R^2(1+\chi_0)}{2Dh^2} \int_0^\pi d\vartheta \frac{(\sin\vartheta)^3}{[4 + 4(R/h)\cos\vartheta + (R/h)^2]^{3/2}} \\
&\stackrel{(R/h \leq 1)}{=} -\frac{b\kappa R^2(1+\chi_0)}{12Dh^2} = -\frac{R^2(1+\chi_0)}{12h^2} U_0 \,\mathrm{sgn}(b).
\end{aligned} \tag{16}$$

At the equilibrium height $h_{eq}$ this contribution balances the free space velocity $U_{fs} = [(1-\chi_0^2)/4]\, U_0 \,\mathrm{sgn}(b)$, so that

$$h_{eq}/R = [3(1-\chi_0)]^{-1/2}. \tag{17}$$

## V. NONUNIFORM SURFACE MOBILITIES

In this section we concisely discuss how small to moderate deviations from the case of uniform mobilities, i.e., $\beta := b_{inert}/b_{cap} \neq 1$ but close to 1, affect the steady states determined at $\beta = 1$. A detailed study of the phenomenology for $\beta \neq 1$, including the case $\beta < 0$, will be presented elsewhere.

### A. Effect on sliding and hovering states

We consider whether the sliding and hovering states previously obtained for $\beta = 1$ persist for $\beta$ deviating from $\beta = 1$. Varying $\beta$ for $\chi_0 = 0.4$, $\chi_0 = 0.5$, and $\chi_0 = 0.6$ we find that, for each degree of coverage, there is a nonzero interval of $\beta \simeq 1$ within which sliding does occur (see Fig. 3). The size and and the location of this interval varies

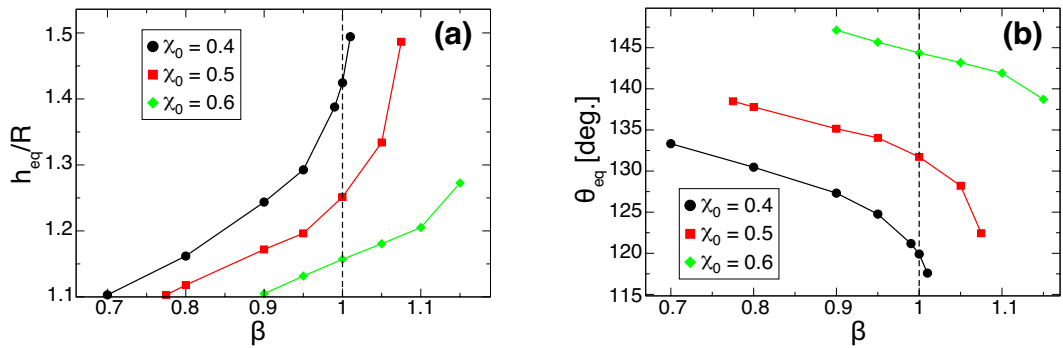

FIG. 3: (a) Variation of the sliding height $h_{eq}/R$ as function of $\beta = b_{inert}/b_{cap}$ for three catalyst coverages $\chi_0$. For each $\chi_0$, there is an interval of $\beta$ with upper and lower bounds for sliding to occur. Slightly beyond the farmost right data points there are no sliding states. It is unclear whether there is a genuine maximum size of this interval, because $h_{eq}/R$ approaches the phase space cutoff $h/R = 1.1$ as $\beta$ decreases. In any case there is a nonvanishing interval of $\beta$, containing $\beta = 1$, within which the sliding states do occur. (b) Variation of the sliding orientation $\theta_{eq}$ as function of $\beta$ for the same three catalyst coverages as in (a).

with coverage. For instance, for $\chi_0 = 0.4$, $\beta = 1.01$ is the largest value of $\beta$ for which sliding occurs. (There is no sliding attractor for $\beta = 1.02$; thus the exact value of the upper bound of the interval lies between $\beta = 1.01$ and $\beta = 1.02$.) On the other hand, for $\chi_0 = 0.6$ sliding occurs for $\beta$ as large as $\beta = 1.15$. As expected, the height $h_{eq}$ and the angle $\theta_{eq}$ for sliding vary as a function of $\beta$. As $\beta$ decreases, $h_{eq}$ decreases. Eventually, $h_{eq}/R$ reaches $h/R = 1.1$, which is the cutoff of the region of phase space we consider. It is therefore uncertain whether there is a genuine lower bound on $\beta$ for sliding.

Concerning the hovering states, we have considered $\chi_0 = 0.85$, $\chi_0 = 0.9$, and $\chi_0 = 0.95$, with $\beta \in [0, 1.5]$. We have not found any genuine bounds for hovering within this interval of $\beta$. However, as for sliding, $h_{eq}$ for hovering depends on $\beta$ such that it decreases upon increasing $\beta$ and finally reaches the cutoff $h/R = 1.1$.

### B. Identification of chemical and hydrodynamic wall effects

In Section III we outlined how to separate the hydrodynamic and chemical contributions to the angular velocity $\dot{\theta}$ of a self-propelled particle. The method consists of employing in the calculation of the translational and angular velocities $\mathbf{U}$ and $\mathbf{\Omega}$ either (i) a prescribed distribution of slip velocities around the particle given by that corresponding to the self-phoresis of the same particle with the same orientation $\theta$ in free space [10], i.e., $\mathbf{v_s}(\mathbf{r}) = \mathbf{v_s}^{fs}(\mathbf{r})$, or (ii) a prescribed hydrodynamic stress tensor given by that corresponding to a sphere moving through an unbounded fluid [5]. Here we have repeated this analysis for half coverage ($\chi_0 = 0$) and unequal cap and inert surface mobilities, i.e., $\beta \neq 1$. Figure 4 shows such results for $\beta = 0.9$.

It is interesting to directly compare Figs. 1 and 4 which correspond to $\beta = 1$ and $\beta = 0.9$, respectively. For these two cases, the hydrodynamic contributions to $\dot{\theta}$, shown in Figs. 1(b) and 4(b), are similar and negative everywhere in the region $0 < \theta < \pi$. However, for $\beta = 0.9$ the chemical contributions to rotation are no longer negligible (which is

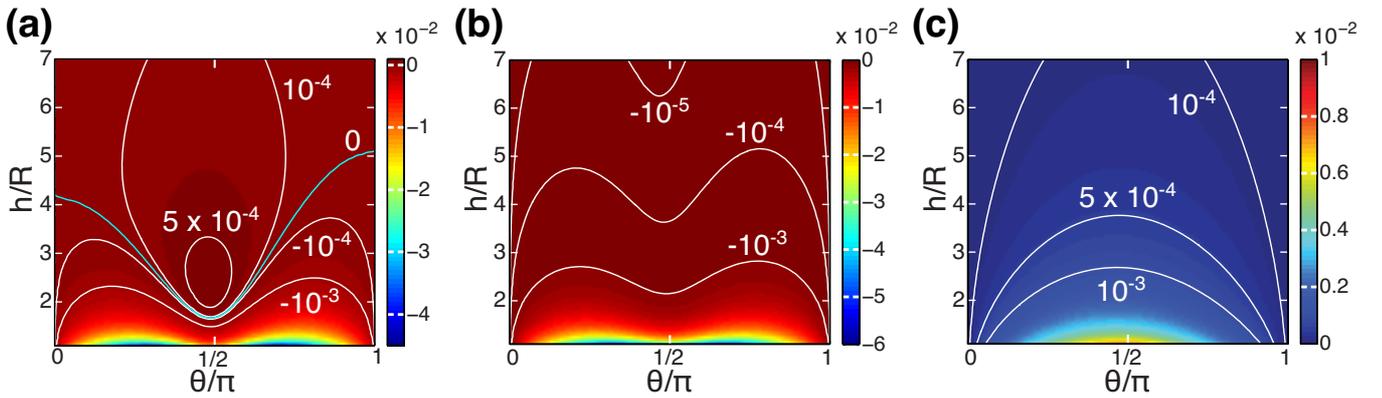

FIG. 4: (a) Angular velocity $\dot{\theta}/\Omega_0$ as a function of height $h/R$ and orientation $\theta$ for half coverage ($\chi_0 = 0$) and unequal surface mobilities: $\beta = b_{inert}/b_{cap} = 0.9$. Throughout, white curves correspond to constant values of $\dot{\theta}$. In contrast to the case $\chi_0 = 0$ and $\beta = 1$ [Fig. 1(a)], here there exists a curve (shown in cyan color) along which $\dot{\theta} = 0$. (b) Angular velocities $\dot{\theta}/\Omega_0$ obtained by using in the reciprocal theorem the free space phoretic slip $\mathbf{v}_s^{fs}$ around the particle, i.e., neglecting the influence of the wall on the number density of solute but including the influence of the wall on the hydrodynamic flow. (c) $\dot{\theta}/\Omega_0$ obtained by using the corresponding free space hydrodynamics stress tensor of a translating or rotating sphere in each of the $j = 1\ldots 6$ dual problems employed in the reciprocal theorem, i.e., neglecting the effect of the wall on the hydrodynamics but including the chemical effect. In strong contrast to the results shown in Fig. 1(c), here due to $\beta \neq 1$ chemical gradients created by the wall significantly drive the rotation of the particle.

in line with the fact that for $\beta \neq 1$ Smoluchowski's result for $b(\mathbf{r}) = const$ [1] does not apply), and they are positive everywhere for $0 < \theta < \pi$ [Fig. 4(c)]. On the cyan curve in Fig. 4(a), the two contributions to rotation are in balance, i.e., $\dot{\theta} = 0$.

As previously discussed, we do not expect Fig. 4(b) and Fig. 4(c) to exactly sum to Fig. 4(a), owing to coupling between chemical and hydrodynamic effects at $\mathcal{O}(h^{-4})$. Since the chemical contribution is no longer negligible for $\beta \neq 1$, we can inspect the accuracy of our approach of decomposing the hydrodynamic and chemical effects by comparing the sum of Fig. 4(b) and Fig. 4(c) with Fig. 4(a). The sum of the two individually isolated contributions is shown in Fig. 5. The resulting reconstructed angular velocity function strongly resembles Fig. 4(a), which includes coupling between chemical and hydrodynamic effects, and was used to integrate the trajectories in Fig. 4 of the main text. We stress that all trajectories and phase plots in the main text include coupling between chemical and hydrodynamic effects.

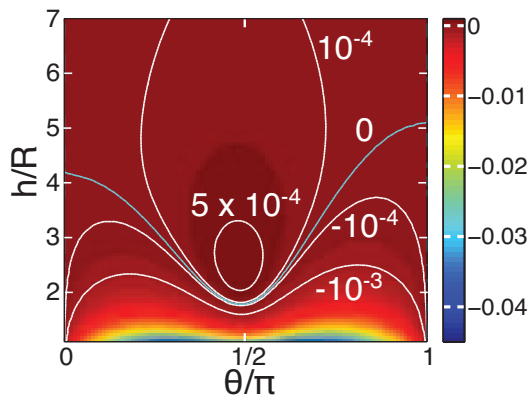

FIG. 5: The sum of Fig. 4(b) and Fig. 4(c), the individually isolated hydrodynamic and chemical contributions to rotation of a particle with $\beta = 0.9$ and $\chi_0 = 0$. The resulting angular velocity function is very similar to Fig. 4(a), the angular velocity function obtained when the two contributions are not separated, i.e. via the numerical method used to obtain the figures in the main text.

[1] J. L. Anderson, Ann. Rev. Fluid Mech. **21**, 61 (1989).

[2] R. Golestanian, T. B. Liverpool, and A. Ajdari, Phys. Rev. Lett. **94**, 220801 (2005).

[3] M. N. Popescu, S. Dietrich, M. Tasinkevych, and J. Ralston, Eur. Phys. J. E **31**, 351 (2010).

[4] C. Pozrikidis, *A Practical Guide to Boundary Element Methods with the Software Library BEMLIB* (CRC Press, Boca Raton, 2002).

[5] J. Happel and H. Brenner, *Low Reynolds Number Hydrodynamics* (Prentice-Hall, Englewood Cliffs, NJ, 1965).

[6] W. M. Deen, *Analysis of Transport Phenomena* (Oxford University Press, New York,, 1998).

[7] J. R. Blake and A. T. Chwang, J. Eng. Math. **8**, 23 (1974).

[8] S. Spangolie and E. Lauga, J. Fluid Mech. **700**, 105 (2012).

[9] K. Ishimoto and E. A. Gaffney, Phys. Rev. E **88**, 062702 (2013).

[10] R. Golestanian, T. B. Liverpool, and A. Ajdari, New J. Phys. **9**, 126 (2007).



\end{document}